\documentclass[a4paper,11pt]{article}
\usepackage{amsmath}
\usepackage{amssymb}
\usepackage{indentfirst}
\usepackage[english]{babel}
\usepackage{graphicx} 
\usepackage{subcaption}
\usepackage[dvipsnames,svgnames,table]{xcolor}
\usepackage{epstopdf}
\usepackage{mathrsfs}
\usepackage{ulem}
\usepackage{enumerate}
\usepackage{enumitem}
\usepackage{amsthm}
\usepackage{cite}
\usepackage{float}
\usepackage{xcolor,soul}
\usepackage{booktabs}
\usepackage{cleveref}

\newcommand{\be}{\begin{equation}}
\newcommand{\ee}{\end{equation}}
\newcommand{\bea}{\begin{eqnarray}}
\newcommand{\eea}{\end{eqnarray}}

\newcommand{\volume}{{\ooalign{\hfil$V$\hfil\cr\kern0.08em--\hfil\cr}}}
\newcommand{\Rey}{{\mathrm{Re}}}

\DeclareMathAlphabet\mathbfcal{OMS}{cmsy}{b}{n}

\def\pmb#1{\setbox0=\hbox{#1}
	\kern-.025em\copy0\kern-\wd0
	\kern.05em\copy0\kern-\wd0
	\kern-.0125em\raise.0433em\box0 } 

\def\bOmega {\pmb {$\Omega$}}
\def\bomega {\pmb {$\omega$}}

\def\bv{\pmb {v}}
\def\bu{\pmb {u}}

\def\bQ{\pmb {Q}}
\def\bn{\pmb {n}}
\def\bx{\pmb {x}}
\def\bbe{\pmb {e}}

\def\bF{\pmb {F}}
\def\bT{\pmb {T}}

\title{Towards Particle-Resolved Accuracy in Euler-Lagrange Simulations of Multiphase Flow Using Machine Learning and Pairwise Interaction Extended Point-particle (PIEP) Approximation}

\author{
	S. Balachandar%
	\thanks{Corresponding Author, Professor, University of Florida, bala1s@ufl.edu}, \,
	W. C. Moore%
	\thanks{Graduate Student, University of Florida},  \, 
    G. Akiki%
	\thanks{Assistant Professor, {Notre Dame University - Louaize, Lebanon}},  \,and
	K. Liu%
	\thanks{Graduate Student, University of Florida}
	\\
	{\normalsize\itshape
		University of Florida, Gainesville, FL, 32611} \\
}

\date{}

\begin{document}
	\maketitle
	
\begin{abstract}
	
	This study presents two different machine learning approaches for the modeling of hydrodynamic force on particles in a particle-laden multiphase flow. Results from particle-resolved direct numerical simulations (PR-DNS) of flow over a random array of stationary particles for eight combinations of particle Reynolds number ($\Rey$) and volume fraction ($\phi$) are used in the development of the models. The first approach follows a two step process. In the first {\it {flow prediction}} step, the perturbation flow due to a particle is obtained as an axisymmetric superposable wake using linear regression. In the second {\it{force prediction}} step, the force on a particle is evaluated in terms of the perturbation flow induced by all its neighbors using the generalized Fax\'en form of the force expression. In the second approach, the force data on all the particles from the PR-DNS simulations is used to develop an artificial neural network (ANN) model for direct prediction of force on a particle. Due to the unavoidable limitation on the number of fully resolved particles in the PR-DNS simulations, direct force prediction with the ANN model tends to over-fit the data and performs poorly in the prediction of test data. In contrast, due to the millions of grid points used in the PR-DNS simulations, accurate flow prediction is possible, which then allows accurate prediction of particle force. This hybridization of multiphase physics and machine learning is particularly important, since it blends the strength of each, and the resulting pairwise interaction extended point-particle (PIEP) model cannot be developed by either physics or machine learning alone.
\end{abstract}

\section{Introduction}	\label{sec:1}
Particle-resolved direct numerical simulations (PR-DNS) of a multiphase flow are fundamental in the sense that they are nearly free of closure assumptions. That is, the governing Navier--Stokes equations for the fluid flow around the particles and the Lagrangian equations of motion for all the particles are faithfully solved. However, this approach requires that the details of the flow around the particles, including their boundary layers and wakes, be fully resolved. This can be accomplished with either a time-dependent grid that is body-fitted around the particles or with an immersed boundary method (IBM), where the interface between the particles and the surrounding fluid are tracked with a set of Lagrangian markers. Both these approaches are computationally expensive since they require the grid to be more than an order of magnitude smaller than the particle size. As a result, even with the worlds largest computers the number of particles that can be considered with the PR-DNS approach is limited to about $O(10^5)$ particles. This number is likely to increase in the coming decades but will remain far smaller than what is needed in many multiphase flow applications.

The alternative that has been widely used is the Euler-Lagrange (EL) approach, where again the fluid phase governing equations are solved in the Eulerian frame, and the particles are tracked using their equations of motion. The primary difference from the PR-DNS approach is that the fluid phase governing equations in the EL approach have been filtered (or averaged) over a length scale much larger than the particle size. All microscale flow features that are below the filter scale (i.e., all flow scales of the order of particle size and smaller) are averaged out and therefore need not be explicitly resolved. Thus, EL approach has the following advantages: (i) The interface between the particles and the fluid need not be resolved either using a body-fitted grid or IBM. (ii) The fluid governing equations can be solved over the entire volume without demarcating the region inside versus outside the particles. (iii) Most importantly, the Eulerian grid can be an order of magnitude or more larger than the particle size. As a result of these computational advantages, EL approach can investigate much larger systems involving billions of particles. 

The single biggest disadvantage of the EL approach is that its accuracy is not guaranteed, unlike the PR-DNS approach. With a rigorous averaging process, an accurate set of filtered governing equations for the EL approach can be obtained \cite{Anderson_jackson, Saffman, cap_desjardin, bala_liu}. The filtered equations will include the following closure terms which must be modeled: (i) the momentum exchange between a particle and the surrounding flow modeled as force and torque exerted on the particle and (ii) the subgrid stress associated with the velocity fluctuations that were averaged in the filtering process. These closure models account for the effect of the filtered microscale fluid motion on the dynamics of the particles as well as on the dynamics of the larger fluid scales that have not been filtered.

We hypothesize the existence of perfect closure models for the force and torque on the particles as well as for the subgrid fluid stress. Here the term ``perfect'' applies to a model when its prediction is fully consistent with a PR-DNS. In other words, the force and torque of the particles predicted with perfect closure models for a given macroscale flow will equal those from a PR-DNS simulation. An EL simulation with perfect closure models can then be termed ``ideal-EL''. Two finer points must be discussed. First, even the ``ideal-EL'' will lack the detailed flow information at the microscale, since such information is subgrid  and is not computed in an EL simulation. Second, due to the chaotic nature of the multiphase flow, the agreement between the EL simulation and the macroscales of the corresponding PR-DNS can only be statistical. Thus, even the ``ideal-EL'' is ideal only in its ability to capture the macroscale structure and the dynamics of the PR-DNS in a statistical sense.  

A major quest of multiphase flow research has thus been the development of force models on particles (and to a lesser extent the development of subgrid stress models) that approach the accuracy of the perfect closure model. Towards this quest, we inquire what such a perfect closure model would entail. In its most general form, the particle force model should depend on (i) the current state and the past history of its motion (i.e., its position as a function of time, from which particle velocity and acceleration can be obtained), (ii) the current state and past history of all other particles, and (iii) the current state and the past history of the filtered macroscale flow field. Clearly, such an elaborate dependency corresponds to an impossibly complex model. Furthermore, such a model will be computationally very expensive and its associated accuracy may not be warranted. Our goal is to improve the accuracy of the model to the extent that quantities of interest can be predicted with confidence at affordable computational expense, so that the improved model can be of practical value. The particle force model that is in common use is the finite-Reynolds number extension of the Maxey-Riley equation \cite{maxey_riley1983, bala_annrev, crowe_etal}. In this model, the force on the particle depends only on the current state and the past history of that particle, and the filtered macroscale flow at the particle location. This model is thus local in nature - it neither depends on the motion of other particles, nor on the fluid flow away from the particle of interest. 

In the Maxey-Riley equation, the force on the particle is separated into quasi-steady, stress-divergence, added-mass, viscous unsteady and lift components. Analytical expressions of each component are available for an isolated particle in the low Reynolds number limit, which are then empirically corrected for finite values of Reynolds number ($\Rey$) and particle volume fraction ($\phi$). Correlations of quasi-steady drag that are dependent on $\Rey$ and $\phi$ are available in the literature \cite{beetstra2007drag, bogner2015drag, tenneti2011drag, zaidi2014new, tang2015new} (we will refer to these as $(\Rey,\phi)$-dependent correlation). Similar analytical expressions of the added-mass coefficient as a function of $\phi$ have also been obtained \cite{sangani}. 

For example, the above referenced $(\Rey,\phi)$-dependent quasi-steady drag correlations provide an excellent approximation of the actual force, in an average sense. By this we mean the following: in a very large system, if we sort all the particles according to their local Reynolds number and particle volume fraction, then the mean force averaged over all the particles within each bin is well predicted by the $(\Rey,\phi)$-dependent quasi-steady drag relations. Two different mechanisms of departure from this $(\Rey,\phi)$-dependent description have been considered recently. First, there is growing recognition that even on an average sense, $(\Rey,\phi)$-dependent correlations are accurate only in case of homogeneous and isotropic distribution of particles. In cases of inhomogeneous particle distributions, $\phi$ alone is not sufficient to describe the local state of particles, the mean drag force will additionally depend on $\nabla \phi$ \cite{su_zhao, rubintein}. In cases of homogeneous, but anisotropic, distribution of particles, the direction of anisotropy will matter and the drag along this direction will be different from that along the transverse directions. Second, even in a statistically homogeneous and isotropic distribution of particles, the force on individual particles substantially deviate from that predicted by the $(\Rey,\phi)$-dependent correlation \cite{akiki2016force}. This deviation depends on the precise manner in which the neighbors are located around the particle with respect to the direction of local macroscopic flow. For example, if one or more of the neighbors are located immediately upstream of the particle, the perturbation flow induced by the upstream neighbors will result in a substantial reduction in the drag force compared to that predicted by the mean correlation. On the other hand, if the neighbors are located such that their perturbations direct an accelerated flow towards the particle, the drag on the particle will be substantially larger. 

In essence, the relative location of the neighboring particles matter greatly in evaluating the force on individual particles and their subsequent motion. It must be emphasized that in an EL simulation, fortunately, each particle has access to the relative position and motion of all its neighbors. Therefore, we seek an improved force model that goes beyond the $(\Rey,\phi)$-dependent mean description to include the specific perturbation effect of neighbors. Such a neighbor-dependent force model may automatically account for the effect of inhomogeneity and anisotropy, since by construction the model now depends on how the neighboring particles are distributed. However, it must be cautioned that even small errors in particle motion can lead to substantial differences in particle distribution at later time. It is thus important to compare the EL simulation results against PR-DNS and experiment results for both single and two-particle statistics.

Results from PR-DNS of flow through a random array of particles were used in developing the $(\Rey,\phi)$-dependent quasi-steady drag relations \cite{beetstra2007drag,bogner2015drag,tenneti2011drag,zaidi2014new,tang2015new}, where a simple curve fit through the quasi-steady drag as a function of  $(\Rey,\phi)$ was sought. In our quest towards the perfect closure model, here we are interested in using the PR-DNS data to develop a more complex force model that systematically includes the effect of neighbors. However, an important challenge arises as we expand the dependence of the model to quantities other than local $\Rey$ and $\phi$. The more complex neighbor-dependent model, in addition to its dependence on $\Rey$ and $\phi$, will also depend on the position, velocity, and acceleration of every one {of} its neighbors. Thus, with the inclusion of each additional neighbor's influence in determining the force on a particle, the number of independent variables in the modeling of force increases by 9. \footnote{In fact, we should also include to this list angular velocity and angular acceleration of each neighbor. This increases the number of independent variables associated with each additional neighbor to 15.} With $N-1$ neighbors taken into account, the force on the particle is determined by the $N$-body configuration. Thus, even for small values of $N$, the number of independent variables dramatically increases and simple curve fitting is not an option.

\subsection{Hybrid Approach: Machine Learning with Multiphase Physics}
We seek the help of machine learning algorithms to develop an improved force model that takes into account the effect of neighbors. The accuracy of machine learning algorithms depends on the quantity of training data provided to the training process. Thus, a naive application of machine learning algorithm will require large amounts of PR-DNS training data of drag forces on particles for a wide range of values of position, velocity and acceleration of neighbors. But such extensive data is beyond our current capability. So we simplify the problem substantially by invoking the pairwise interaction approximation, where the $N$-body problem is simplified as $(N-1)$ pairwise interactions. Then, we proceed to exploit our knowledge of multiphase flow physics and simplify the problem with the following two step process: (i) {\underline {Flow prediction:}} Machine learning, in conjunction with the PR-DNS data of steady flow over a random array of frozen particles, is first used to approximate the perturbation flow due to a neighbor. Superposition of the perturbation flow due to all the particles within a system allows accurate prediction of the microscale flow just from the position and motion of all the particles. (ii) {\underline {Force Prediction:}} The next step involves calculation of the perturbation force on a particle from each of its neighbors' perturbation flow taken one at a time, using the generalized Fax\'en form given by the Maxey--Riley equation \cite{maxey_riley1983, annamalai_bala}. 

The above steps of {\underline {flow prediction}} followed by {\underline {force prediction}} is crucial for a successful implementation of the machine learning algorithm. The PR-DNS data available to us for training the machine learning algorithm typically consists of force and torque information on a few thousand particles, however the flow field around these particles is resolved using tens of millions of grid points. Thus, far more training data is available for accurate prediction of the perturbation flow. Direct force prediction from PR-DNS data is not nearly as accurate, due to the far fewer force/toque training data. In fact, the need for the two step prediction process is even more compelling. Flow prediction through machine learning will be in the context of a stationary random array of particles. But the subsequent application of force prediction using Generalized Fax\'en law will be in the context of freely moving distribution of particles. For otherwise, a direct force prediction through machine learning (without the intermediate flow prediction step) will require far more PR-DNS training data that must be obtained in the context of freely moving particles. Thus, our strategy is to use machine learning algorithms for accurate flow prediction followed by multiphase theory for dynamic force prediction. The above hybridization of multiphase physics and machine learning is particularly important, since it blends the strength of each, and the resulting force model cannot be developed by either physics or machine learning alone. While the physics provides the decomposition of force into components and their functional forms, machine learning extracts the detailed perturbation influence of the neighbor. Another advantage of flow prediction should be noted - the predicted flow at the microscale will allow us to calculate other closure terms of the EL governing equations, such as subgrid stress and subgrid heat flux (see \cite{Moore-bala2019}).

In this paper we will summarize and extend the recent efforts in the development of pairwise interaction extended point-particle (PIEP) model, which systematically accounts for the perturbative effect of neighboring particles on top of the ($\Rey$,$\phi$)-dependent quasi-steady drag relation \cite{akiki2017pairwise, akiki2017pairwiseJCP, moore2019hybrid}. Section 2 presents the limitations of the $(\Rey,\phi)$-dependent mean force model and make the case for incorporating the effect of neighbors. In section 3 we describe flow prediction as the summation of superposable wake contribution from all the particles within the system and in section 4 we present a machine leaning approach for obtaining the perturbation flow of a particle as the superposable wake. Section 5 describes the step towards force prediction, where the predicted superposable wake is used to calculate maps of force influence on neighbors. In section 6 force prediction is discussed in the context of a stationary array of particles. Here the predictive capability of the present PIEP model, where force prediction is enabled after flow prediction, is compared against the results of an artificial neural network (ANN) model where force on particles is directly predicted after being trained with the PR-DNS data. Section 7 discusses the predictive capability of the PIEP model in the dynamic context of freely settling distribution of particles. In section 8 PIEP torque model is presented followed by conclusions in section 9.

\section{Limitations of $(\Rey,\phi)$-dependent Mean Force Model} \label{sec:2}
In this section we will highlight the strengths and weaknesses of the standard $(\Rey,\phi)$-dependent mean force model. By focusing on how this model deviates from the so called ``perfect model'', we will try to devise improvements in the subsequent sections. The PR-DNS simulations whose results will be used in this study have been discussed in \cite{akiki2016immersed,akiki2016force}. These fully resolved simulations consider flow through a cubic domain within which a monodisperse array of stationary spheres are randomly distributed with uniform probability. The diameter of the particle ($d_*$) is chosen as the length scale and therefore the non-dimensional diameter of each particle is set to unity, and the non-dimensional volume of the cubic domain is $(3\pi)^3$. The number of particles within the domain determines the particle volume fraction within the cubic domain. The boundary condition along the flow direction ($x$) is chosen to be periodic. Along the normal $y$-direction the boundary condition is periodic, and no-stress boundary conditions are used in the $z$-direction. As discussed in \cite{akiki2016force}, away from the no-stress boundaries the macroscale pressure gradient remains statistically uniform and all further results to be presented will be limited to this inner region. The mean volume fraction of particles in this region is $\phi$ and the average streamwise fluid velocity is $\langle u_{*} \rangle$.  An immersed boundary method is used to fully resolve the
flow around the particles. The two parameters of the PR-DNS simulations are the Reynolds number $ \Rey  = \langle u_{*} \rangle d_* /\nu_*$ and the mean particle volume fraction, where $\nu_*$ is the kinematic viscosity of the fluid. 

Results from eight different $ \Rey $ and $ \phi $ combinations will be discussed, which are listed in \Cref{tab:DNScases}. Each case is simulated several times with each realization consisting of a different random distribution of particles of uniform probability. The number of realizations {($NOR$)} and the total number of particles to be investigated in each case {($N_{pr}$)} is also listed in \Cref{tab:DNScases}. The flow remains steady in all cases except {cases 3, 6, and 9, which shows pseudo-turbulent fluctuations} due to the Reynolds number being higher than the critical value for onset of unsteadiness. In these cases the results presented are averaged over time.

\begin{table}
	\begin{center}
		\begingroup
		\setlength{\tabcolsep}{30pt} 
		\renewcommand{\arraystretch}{1.2} 
		\begin{tabular}{||c c c c c||} 
			\hline
			Case & $\phi$ & $Re$ & $NOR$ & $N_{pr}$ \\ [0.5ex] 
			\hline\hline
			1 & 0.11 & 39 & 10 & 108 \\ 
			\hline
			2 & 0.11 & 70 & 10 & 108 \\ 
			\hline
			3 & 0.11 & 173 & 10 & 108 \\ 
			\hline
			4 & 0.21 & 16 & 8 & 214 \\ 
			\hline
			5 & 0.21 & 86 & 7 & 214 \\ 
			\hline
			6 & 0.45 & 3 & 8 & 459 \\ 
			\hline
			7 & 0.45 & 21 & 8 & 459 \\ 
			\hline
			8 & 0.45 & 115 & 8 & 459 \\ 
			\hline
		\end{tabular}
	\endgroup
		\caption{{Parameters of the PRDNS cases.}}
		\label{tab:DNScases}
	\end{center}
\end{table}

\begin{figure}[H]		   	
		\begin{center}
			\includegraphics[width=0.35\linewidth]{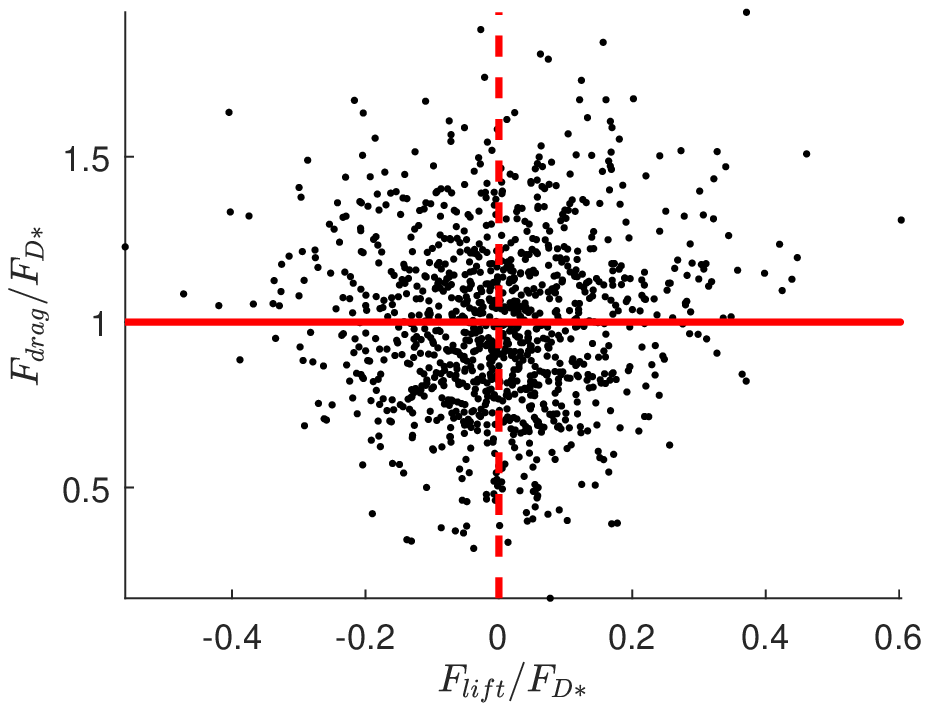}
			\includegraphics[width=0.35\linewidth]{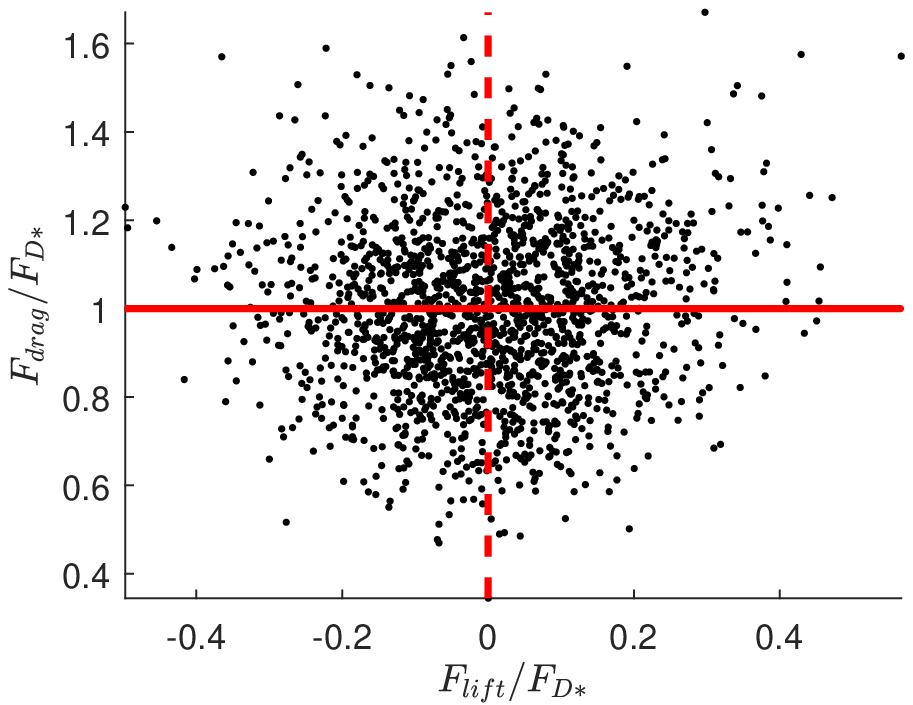}
			\includegraphics[width=0.35\linewidth]{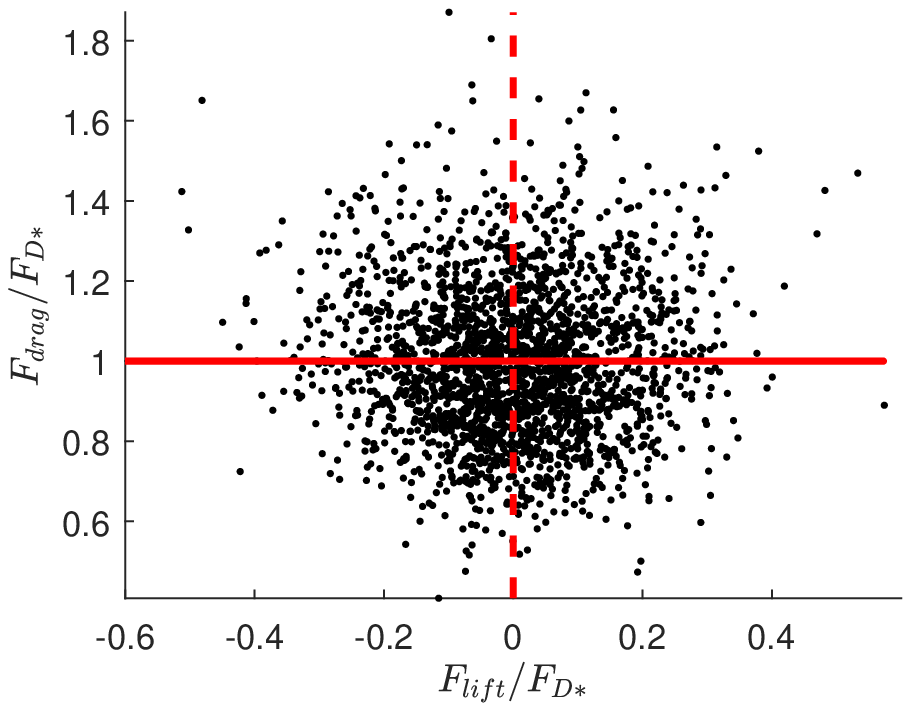}
		\end{center}   
	\caption{{Scatter plot of normalized drag versus normalized lift force obtained from PR-DNS for (a) $ \Rey = 39$ and $ \phi  = 0.11$, (b) $ \Rey  = 16$ and $ \phi  = 0.21$, (c) $ \Rey  = 21$ and $ \phi  = 0.45$.}}
	\label{dragVslift}
\end{figure}

The normalized drag and lift forces on the particles are shown in \Cref{dragVslift} for three of the eight cases, with the results being qualitatively similar for the other cases. Each symbol of the scatter plot corresponds to normalized drag force plotted against the normalized lift force for a particle within the random array. In each case, the results from all the realizations are plotted. The normalization is with respect to drag force prediction using the ($\Rey$,$\phi$)-dependent correlation given in \cite{tenneti2011drag} as
\begin{equation}
 F_{D*} = F_{st*} (1-\phi) \left[ \frac{1+0.15 \Rey^{0.687}}{(1-\phi)^3} + \frac{5.81 \, \phi}{(1-\phi)^3}+
 \frac{0.48 \, \phi^{1/3}}{(1-\phi)^4}+\phi^3 \, Re \left( 0.95 + \frac{0.61 \, \phi^3}{(1-\phi)^2} \right)
 \right] \, , \label{eq:tenneti} 
\end{equation}
where $F_{D*}$ is the dimensional drag force on the particle, and it is written as the corresponding Stokes drag on an isolated particle, $F_{st*}$, multiplied by the ($\Rey$,$\phi$)-dependent correction factor. In applying the above expression to normalize the drag and lift forces on the particles within the array, we take $\Rey$ to be the Reynolds number based on the macroscale flow and $\phi$ to be the mean volume fraction within the array. Since these two parameters are the same for every particle within the random array, the normalization remains the same for all particles in each case. Also plotted is an horizontal line corresponding to a normalized drag of unit magnitude. It can be observed from the scatter plot that the mean value of normalized drag obtained from PR-DNS is close to unity and thus the above ($\Rey$,$\phi$)-dependent correlation provides a good approximation of the mean drag in all cases considered. 

However, the spread of the actual normalized drag force is quite dramatic. For example, in case 1 ($ \phi = 0.11$, $ \Rey  = 39$) the lowest and the highest drag are {$0.16$ and $1.94$}, which are substantially different from the normalized mean value of unity. From the distribution of drag values it is clear that the smallest and the largest drags are not simply statistical outliers. There is consistent and systematic departure from the mean value. In an EL simulation, if the drag on the particles were to be calculated based on the above ($\Rey$,$\phi$)-dependent correlation, substantial error will incur in the drag force of individual particles, which in turn will result in incorrect evaluation of particle motion and incorrect feedback of momentum to the fluid. The behavior is similar for the other two cases shown in \Cref{dragVslift} and also in cases not shown. As mentioned earlier, the substantial increase in drag above the mean value is due to the channeling of flow towards the particle by upstream neighbors, and the substantial decrease is due to the blocking of the flow by neighbors located directly upstream. It is clear from the picture that ($\Rey$,$\phi$)-dependent drag is not sufficient to describe the particle-to-particle variation in the drag force seen within the random array. 

The horizontal axis of \Cref{dragVslift} shows the scatter in the normalized transverse force on the particle. In the PR-DNS, where the mean flow is along the $x$-direction, the direction of transverse force is on the $y-z$ plane and its precise orientation varies from particle to particle. It must be pointed out that given the Reynolds number $\Rey$ and the mean volume fraction $\phi$, without the additional knowledge of where the neighbors are located, the best estimate of lift force is zero {(shown as the vertical line in} \Cref{dragVslift} {for reference)}. The fact that many of the particles have substantial transverse force is due to the specific arrangement of their neighbors. The non-zero magnitude of the transverse force is substantial and often comparable to the drag force and thus cannot be ignored. Thus, evaluating the drag and transverse forces based on only $\Rey$ and $\phi$ will tend to substantially under-estimate the level of particle-to-particle variation. As a result, in an EL simulation of multiphase flow where the particles are allowed to freely move, the dispersion of particles will be underestimated if the particle force is computed only based on ($\Rey$,$\phi$)-dependence. Similar substantial particle-to-particle variation exists in other quantities such as torque on the particle, heat and mass transfer from the particles and have been highlighted in \cite{akiki2017pairwise,akiki2017pairwiseJCP,moore2019hybrid, he_tafti, he_tafti2, he_tafti3, climent}. In what follows we will present the pairwise interaction extended point-particle (PIEP) model that attempts to incorporate the effect of neighbors.

\section{Flow Prediction Through Superposition} \label{sec:3}
As the first of the two step process outlined in the introduction, in this section we will start with the theoretical framework for accurate prediction of the microscale flow around a random array of particles. From this ability to predict the flow at the microscale, we will then be able to accurately calculate the force on the particles in the following sections. 

Towards this goal, we divide the steady flow field around a random distribution of stationary spherical particles into a macroscale component and a microscale component as
\begin{equation} \label{eq:2}
\bu(\bx) = \bu_0(\bx)+ \sum_{j=1}^{N} \bu'_j (\bx) \, ,
\end{equation}
where the macroscale flow $\bu_0$ is the large scale component, and it does not include the microscale details arising from the boundary layers and wakes around the particles. One way to obtain $\bu_0$ is through a convolution integral of the flow field $\bu$ with a Gaussian filter function \cite{cap_desjardin, ireland_desjardins, bala_liu, liu_bala}. The convolution can be considered as a weighted spatial average of $\bu$ with a specified filter function of length scale $\mathcal{L}$, which is chosen to be much larger than the particle diameter. All the fluid motion at the microscale is due to the perturbing presence of the particle, and are not included in $\bu_0$. They are taken into account in the microscale component $\bu'$. As pointed out in \cite{Moore-bala2019}, the microscale flow can be expressed as a sum of perturbation flow due to each particle in the system. Due to the $N$-body nature of the problem, the perturbation flow due to the $j^{th}$ particle depends on the presence of all other particles in the system. As a result, each $\bu'_j$ depends on a long list of variables that define the $N$-particle configuration (i.e., $(N-1)$ vectors that define the relative position of all the neighbors). As addressed in the introduction, we limit the flow prediction to a static system of stationary particles, and therefore the time variable is absent in \eqref{eq:2}. Generalization of {the} above separation of macro and micro-components to a system of freely moving particles will be presented later in the context of force prediction. Note that in a system of $N$ moving particles, each $\bu'_j$  will additionally depend on the velocity and acceleration of all the particles, thus substantially adding to the already formidable challenge.

To simplify the problem we will make an important approximation that the perturbation flow at a point $\bx$ due to the $j^{th}$ particle depends only on the separation distance $\bx - \bx_j$ from the center of the $j^{th}$ particle and on the properties of the macroscale flow at the $j^{th}$ particle. This approximation has the advantage of rendering the perturbation flow due to the $j^{th}$ particle independent of the presence of all other particles and thereby avoid the difficulty of the $N$-body problem. 

In the present context of steady flow through a cubic box of randomly distributed particles, we take the macroscale flow to be the box average over the region occupied by the fluid. Since $\langle u_{*} \rangle$ is chosen as the velocity scale, in non-dimensional term, $\bu_0(\bx) = \bbe_x$, where $\bbe_x$ is the unit vector along the mean flow direction. In calculating the perturbation flow due to the particles, we take all the particles within the array to be subjected to the same constant macroscale flow of unit non-dimensional magnitude, directed in the $x$-direction, and the volume fraction to be $\phi$. Thus, the perturbation flow induced by each particle is taken to be the same, but centered around each particle. Since we approximate the perturbation flow of each particle to be independent of all other particles, it is axisymmetric about the macroscale flow direction. The perturbation flow due to the $j^{th}$ particle can then be written as
\begin{equation} \label{eq:3}
\bu'_j(\bx) \approx \bu_{sw} (r_j, \theta_j \,|\, \Rey , \phi) \, ,
\end{equation}
where $\bu_{sw}$ (without the subscript ``$j$'') is the universal perturbation flow due a particle that only depends on the Reynolds number $\Rey$ and the volume fraction $\phi$. The perturbation flow $\bu_{sw}$ is axisymmetric about the $x$-axis passing through the center of the particle and thus is a function of the distance $r_j = |\bx - \bx_j|$ and the angle $\theta_j = \cos^{-1}((\bx-\bx_j) \cdot \bbe_x/r_j)$ between the separation vector and the $x$-axis. It must be noted that the axisymmetric perturbation flow has only the radial component $u_{r,sw}$ and the circumferential component $u_{\theta,sw}$ (i.e., the azimuthal component $u_{\varphi,sw} \equiv 0 $). Substituting the above approximations into \eqref{eq:2}, the steady flow through a random array of stationary particles can be approximated as
\begin{equation} \label{eq:4}
\bu(\bx) \approx \bbe_x + \sum_{j=1}^{N} \bu_{sw} (r_j, \theta_j) \, ,
\end{equation}
where the dependence of $\bu_{sw}$ on $\Rey$ and $\phi$ has been suppressed.

\section{Machine Learning for Superposable Wake}
The next step of our quest is to obtain the steady axisymmetric perturbation flow field $\bu_{sw}(r,\theta)$ for varying values of $\Rey$ and $\phi$. This will be accomplished by requiring that the superposition given in \eqref{eq:4} provide the best possible approximation to the steady flow fields obtained in the PR-DNS cases listed in \Cref{tab:DNScases}. This is an {\em inverse problem} and we obtain $\bu_{sw}$ by minimizing the following mean square error
\begin{equation} \label{eq:5}
E^2 = \dfrac{1}{\mathcal{V} M} \sum_{m = 1}^{M} \left [ \int_{\mathcal{V}} |\bu_{DNS,m}(\bx) - \left[ \bbe_x + \sum_{j=1}^{N} \bu_{sw} (r_j, \theta_j) \right] |^2 \, d\mathcal{V} \right] \, ,
\end{equation}
where $\mathcal{V}$ is the volume occupied by the fluid within the cubic box. Here $\bu_{DNS,m}(\bx)$ is the flow field obtained in the $m^{th}$ realization of PR-DNS. In the above $r_j$ and $\theta_j$ are the radial and circumferential components of the separation vector $\bx - \bx_{j,m}$, where $\bx_{j,m}$ is the center of the $j^{th}$ particle in the $m^{th}$ realization. Thus, the term within the inner square parenthesis is the total flow due to all the $N$ particles within the cubic box in the $m^{th}$ realization. 

We now face the inverse problem of finding the steady axisymmetric perturbation flow $\bu_{sw}(r,\theta)$ that will minimize the above defined mean square error. The resulting optimal perturbation flow is termed the {\it {superposable wake}} \cite{Moore-bala2019}, since when superposed as given in \eqref{eq:4} it yields the best approximation to the PR-DNS results in the mean square sense. Even after optimization we expect the error $E$  to remain non-zero, and the error is mainly due to the pairwise approximation that the perturbation flow due to the $j^{th}$ particle is independent of all other particles. A subtle point must be stressed here. Since the perturbation flow is taken to be a function of the volume fraction $\phi$, $\bu_{sw}$ does account for the collective influence of all the other particles. However, it does not account for their specific arrangement. 

Moore \& Balachandar \cite{Moore-bala2019} pursued error minimization using spherical harmonic expansion. Due to the axisymmetric nature of the perturbation flow about the $x$-axis, the perturbation velocity has only non-zero radial and circumferential velocity components that depend on $r$ and $\theta$. The axisymmetric spherical harmonic expansions for the radial and circumferential velocity components are
\begin{equation} \label{eq:6}
u_{r,sw}(r, \theta ) = \sum_{l = 0}^{L} \frac{-f_l(r)}{l(l+1)} \left(
\frac{\partial^2 Y_{l,0}(\theta)}{\partial\theta^2} +\frac{\partial Y_{l,0}(\theta)}{\partial\theta}\frac{\cos\theta}{\sin\theta} \right),
\end{equation}
\begin{equation}\label{eq:7}
u_{\theta,sw}(r, \theta) = \sum_{l = 0}^{L} \frac{1}{l(l+1)}\frac{\partial Y_{l,0}(\theta)} {\partial\theta} \left(2 \, f_l(r) + r \, \frac{\partial f_l(r)}{\partial r}
\right) \, ,
\end{equation}				
where $Y_{l,0}$ are the associated Legendre polynomials (see \cite{Moore-bala2019}). The radial function $f_l(r)$ is expanded in the radial direction in terms of the Bessel $(J_l)$ and Neumann $(N_l)$ functions as 
\begin{equation}\label{eq:8}
f_l(r) = \sum_{m = 1}^{M}  \left[ \textrm{a}_{l,m} J_l(k_{l,m} \, r)  + \textrm{b}_{l,m} N_l(q_{l,m} \, r)\right] \, .
\end{equation}	
In the above, $\textrm{a}_{l,m}$ and $\textrm{b}_{l,m}$ are the expansion coefficients that need to be determined for $l = 0, \cdots, L$ and $m = 1, \dots M$. Here $L$ and $M$ are the upper limits of the circumferential and radial modes being considered, whose optimal values are determined as part of the regression analysis. A more accurate representation of the perturbation velocity field requires larger values of $L$ and $M$. But with increasing $L$ and $M$ the number of expansion coefficients $\textrm{a}_{l,m}$ and $\textrm{b}_{l,m}$ to be optimized increases. We note that with the above expansion, the perturbation velocity identically satisfies the incompressibility condition. The radial wave numbers $k_{l,m}$ and $q_{l,m}$ are selected from the range $10^{-1}$ to $10^1$.

Since the expansion coefficients are linear, the flow predicted by the superposable wakes can be reformulated into a linear system of equations. Linear regression can then be used to minimize the error and obtain the optimal values of the expansion coefficients $\textrm{a}_{l,m}$ and $\textrm{b}_{l,m}$. After obtaining the optimal expansion coefficients, the corresponding perturbation velocity field is obtained by carrying out the summations given in \eqref{eq:6} to \eqref{eq:8} \cite{Moore-bala2019}.

\begin{figure}[H]		   	
	\begin{center}
		\includegraphics[width = 6.7in]{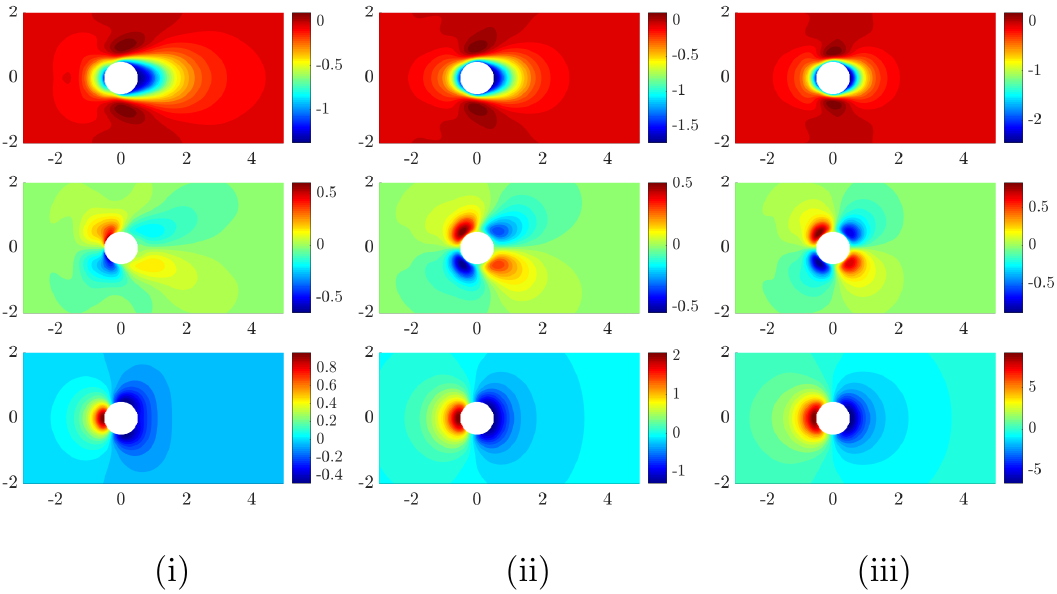}
	\end{center}   
	\caption{The superposable streamwise velocity (top row), transverse velocity (middle row) and pressure (bottom row) fields obtained from regression. The three columns correspond to the three cases: (i) $\Rey = 39$, $\phi = 0.11$, (ii) $\Rey = 16$, $\phi = 0.21$, and (iii) $\Rey = 21$, $\phi = 0.45$.}
	\label{fig2}
\end{figure}

The resulting velocity fields of the superposable wake for three of the cases listed in \Cref{tab:DNScases} are shown in Figure \ref{fig2}. Frames (a) to (c) present contours of perturbation streamwise velocity  (i.e., $u_{x,sw} = u_{r,sw} \cos\theta - u_{\theta,sw}\sin\theta$) and frames (d) to (f) present contours of perturbation in-plane transverse velocity  (i.e., $u_{y,sw} = u_{r,sw} \sin\theta + u_{\theta,sw}\cos\theta$). The superposable velocity and pressure fields presented in Figure \ref{fig2} can be interpretted in teh following way. They represent the perturbation flow created by a particle and they only depend on the Reynolds number and volume fraction at that particle. In a random distribution of particles, each particle within the box will contribute its perturbation flow. This when summed over all the particles and added to the mean macroscale flow, as given in \eqref{eq:2}, will yield the best approximation to the PR-DNS result. Thus, for a given distribution of particles subjected to a uniform macroscale flow, the complex flow field obtained from the summation given in \eqref{eq:2} depends only on the distribution of particles. This synthetic flow well approximates the actual PR-DNS flow and the comparison has been presented in \cite{Moore-bala2019}.

At lower volume fraction, the perturbation velocity is similar to that of an isolated particle of zero volume fraction. At the higher volume fraction, the velocity contours  appear to become fore-aft symmetric, even at finite Reynolds numbers. This is clearly due to the fact that the particle wakes are broken up by the neighbors located downstream. Following \eqref{eq:4} we define the following decomposition of the pressure field
\begin{equation} \label{eq:9}
p(\bx) = p_0(\bx) + \sum_{j=1}^{N} p_{sw} (r_j, \theta_j) \, ,
\end{equation} 
where we have approximated the perturbation pressure due to the $j^{th}$ particle to be independent of the precise location of all other particles and define an axisymmetric superposable perturbation pressure field. The regression algorithm with spherical harmonics is replicated to obtain the superposable pressure field $p_{sw}(r,\theta)$ for varying values of $\Rey$ and $\phi$. These superposable pressure fields are also shown in Figure \ref{fig2}. A complete discussion of the properties of the superposable wakes has been presented in \cite{Moore-bala2019}.

\section{Influence Maps Obtained From Superposable Wake}
\begin{figure}[H]		   	
	\begin{center}
		\includegraphics[width = 2.7in]{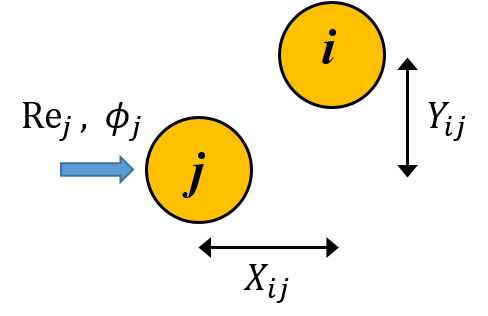}
	\end{center}   
	\caption{A schematic of the perturbation effect of the $j^th$ neighbor whose macroscale Reynolds number and local volume fraction are $(\Rey_j, \phi_j)$ on the $i^{th}$ particle, which is $(X_{ij},Y_{ij})$ away along the flow and transverse directions.}
	\label{fig3}
\end{figure}

The next step towards our goal of calculating the perturbation force on a particle by its neighbor is to translate the superposable wakes of the previous section into influence maps that can be used in force calculation. Consider the $j^{th}$ particle of a random array subjected to a macroscale flow of Reynolds number $\Rey$ surrounded by other particles at an average particle volume fraction of $\phi$. The perturbation flow due to a particle is depicted in Figure \ref{fig2} and it depends on the macroscale flow characterized by the Reynolds number $\Rey$ and the volume fraction $\phi$. We now consider the influence of this perturbation flow on a second particle denoted as the $i^{th}$ particle. The presence of the $j^{th}$ and the $i^{th}$ particle are shown in the schematic of Figure \ref{fig3}, where the macroscale flow at the $j^{th}$ particle is also denoted. Distance from the $j^{th}$ to the $i^{th}$ particle along the flow direction is denoted as $X_{ij}$ and the distance between the particle centers in the normal-to-flow direction is denoted as $Y_{ij}$.

As discussed in \cite{akiki2017pairwise}, we now define the following {\it{perturbation maps}} that measure the influence of the $j^{th}$ particle on its neighbor, the $i^{th}$ particle:
\begin{enumerate}[label=(\roman*)]
\item The surface averaged streamwise velocity perturbation of the $j^{th}$ particle on the $i^{th}$ particle is defined as
	\begin{equation}\label{eq:10}
	\mathscr{M}_{f1}(X_{ij},Y_{ij} \,|\, \Rey,\phi) = \dfrac{1}{\pi} \int_S \bu_{sw}(r,\theta \,|\, \Rey,\phi) \cdot \bbe_x \, dS \, ,
	\end{equation}
where the integral is over the surface of the $i^{th}$ particle, whose center is at a distance $(X_{ij},Y_{ij})$ from the center of the $j^{th}$ particle. The scaling factor $\pi$ is from the surface area of a sphere of unit diameter. The surface average of the superposable wake flow also depends on the Reynolds number and the volume fraction and they should be those of the $j^{th}$ particle, which is the perturbing particle, whose perturbation flow is being considered. Note that in the previous section, the perturbation flow due to the $j^{th}$ particle has been computed taking into account only the collective effect of all the neighbors. Thus, $\bu_{sw}$ is well defined everywhere outside of the $j^{th}$ particle, including the region occupied by the $i^{th}$ particle. Thus, the surface integral of $\bu_{sw}$ is around the surface of a {virtual} sphere centered around $\bx_i$. We take the non-dimensional diameter of the $i^{th}$ particle to be unity, corresponding to a monodisperse system. Otherwise, the surface average will also depend on the non-dimensional diameter of the $i^{th}$ particle.
\item The surface averaged transverse velocity perturbation of the $j^{th}$ particle on the $i^{th}$ particle is defined as
\begin{equation}\label{eq:11}
\mathscr{M}_{f2}(X_{ij},Y_{ij} \,|\, \Rey,\phi) = \dfrac{1}{\pi} \int_S \bu_{sw}(r,\theta\,|\,\Rey,\phi) \cdot \bbe_y \, dS \,.
\end{equation}
This component lies on the plane containing the centers of the two particles and the unit vector $\bbe_x$ passing through the $j^{th}$ particle and is normal to $\bbe_x$.
\item The streamwise component of the surface averaged tractional force around the $i^{th}$ particle due to the perturbation flow of the $j^{th}$ particle can be written as the following volume average 
\begin{equation}\label{eq:12}
\mathscr{M}_{f3}(X_{ij},Y_{ij} \,|\, \Rey,\phi) = \dfrac{6}{\pi}  \int_V \left[ -\nabla p_{sw}(r,\theta\,|\,\Rey,\phi) + \dfrac{1}{\Rey} \nabla^2 \bu_{sw}(r,\theta\,|\,\Rey,\phi) \right] \cdot \bbe_x \, dV \,,
\end{equation}
where the integral is over the volume occupied by the $i^{th}$ particle.
\item Transverse component of the surface averaged tractional force around the $i^{th}$ particle due to the perturbation flow of the $j^{th}$ particle, written as the following volume average 
\begin{equation}\label{eq:13}
\mathscr{M}_{f4}(X_{ij},Y_{ij} \,|\, \Rey,\phi) = \dfrac{6}{\pi} \int_V \left[ -\nabla p_{sw}(r,\theta\,|\,\Rey,\phi) + \dfrac{1}{\Rey} \nabla^2 \bu_{sw}(r,\theta\,|\,\Rey,\phi) \right] \cdot \bbe_y \, dV \,.
\end{equation}
\item Out-of-plane component of the volume averaged perturbation vorticity around the $i^{th}$ particle  
\begin{equation}\label{eq:14}
\mathscr{M}_{f5}(X_{ij},Y_{ij} \,|\, \Rey,\phi) = \dfrac{6}{\pi}  \int_V \left[ \nabla \times \bu_{sw}(r,\theta\,|\,\Rey,\phi) \right] \cdot \bbe_z \, dV \,,
\end{equation}
where $\bbe_z$ is the unit vector in the $z$-direction.
\end{enumerate} 

\begin{figure}[H]		   	
	\begin{center}
		\includegraphics[width = 5.7in]{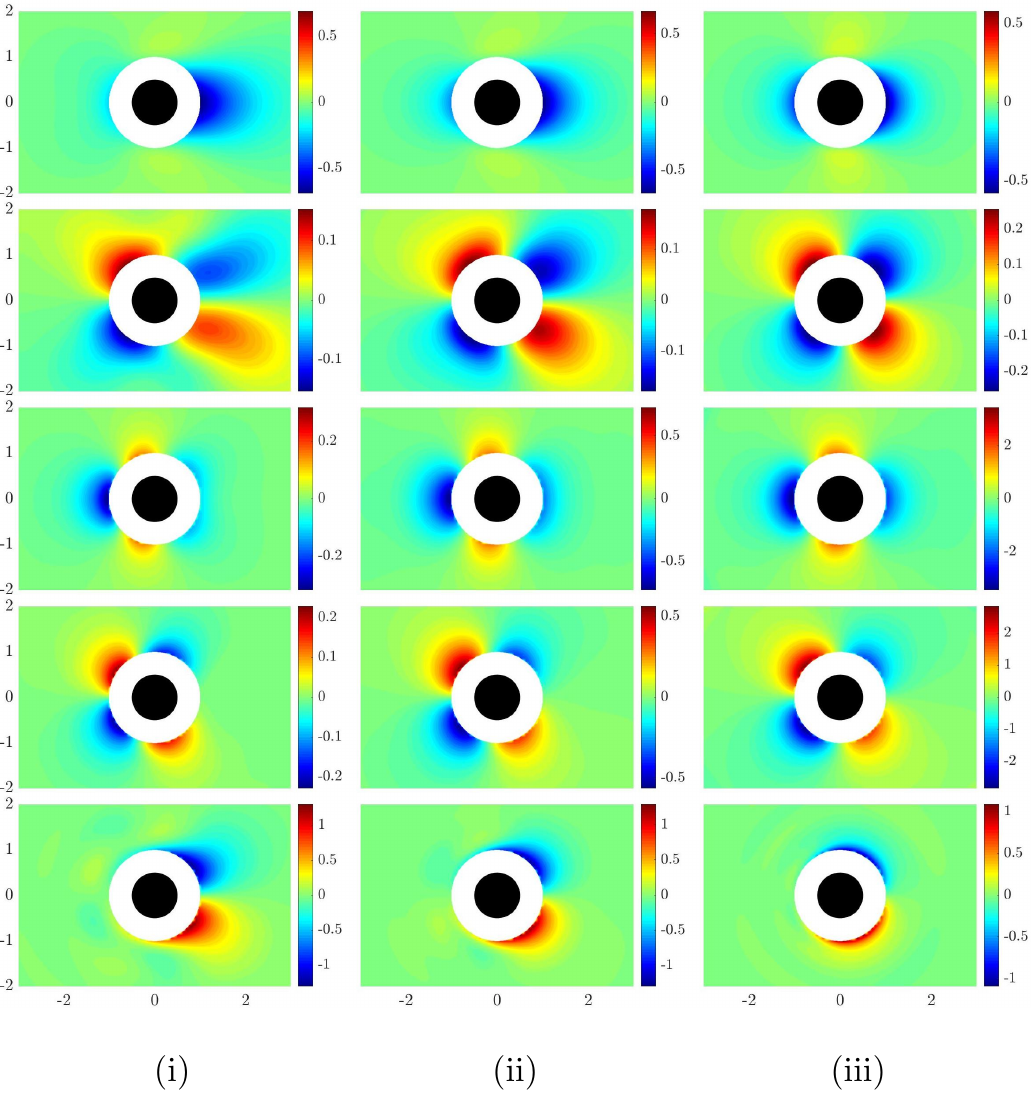}
	\end{center}   
	\caption{Rows corresponding to maps $\mathscr{M}_{f1}$ through $\mathscr{M}_{f5}$. The three columns correspond to the three cases: (i) $\Rey = 39$, $\phi = 0.11$, (ii) $\Rey = 16$, $\phi = 0.21$, and (iii) $\Rey = 21$, $\phi = 0.45$.}
	\label{fig4}
\end{figure}

Plots of these five maps for the three cases shown in Figure \ref{fig2} are presented in Figure \ref{fig4}. The columns correspond to the different cases, while the five rows correspond to maps of $\mathscr{M}_{f1}$ to $\mathscr{M}_{f5}$. These perturbation maps are qualitatively similar to those presented in \cite{akiki2017pairwise}, but they are far more accurate at finite volume fraction due to much improved flow prediction. In \cite{akiki2017pairwise}, the five perturbation maps were generated based on the axisymmetric perturbation flow of an isolated particle and thus was only a function of $\Rey$. These isolated particle's perturbation maps are therefore appropriate only in the limit of low volume fraction. In Figure \ref{fig4}, the left column of perturbation maps corresponding to $\phi = 0.11$ are similar to those presented in \cite{akiki2017pairwise}. The importance of superposable wake is that it provides an accurate representation of the perturbation flow taking into account the collective influence of the neighbors. 

These contour maps must be interpreted in the following manner: the black circle at the center of each frame corresponds to the perturbing $j^{th}$ particle and at any point outside of it, the contour value corresponds to the perturbing influence (streamwise velocity perturbation in the case of the top row) surface or volume averaged over the $i^{th}$ particle, that is centered about that point. The white annular band around the black circle is the excluded region of a monodisperse system, since the center of the $i^{th}$ particle cannot lie in this region. As can be expected, the streamwise velocity perturbation is negative both upstream and downstream of the $j^{th}$ particle, but is weakly positive on the sides of the particle. The cross-stream velocity is such that the $i^{th}$ neighbor is pushed away from the streamwise centerline on the upstream side, while pushed towards the streamwise centerline on the downstream side. The streamwise and cross-stream components of the surface averaged tractional force also follow similar qualitative behavior. The torque effect is primarily limited to neighbors that are on the downstream side. An important observation of significance is that all the perturbation maps decay sufficiently rapidly that the perturbation influence of the $j^{th}$ particle is practically zero when a neighbor is three or four diameters away. 

A systematic change in the perturbation maps with increasing volume fraction is observed, going from the left to the right column of frames. In all these cases, the flow is from left to the right. At the low volume fraction of $\phi =0.11$, the streamwise velocity perturbation map (top row), the streamwise tractional force map (third row), and the torque map (last row) are quite asymmetric about the vertical centerline indicating the strong wake effect that distinguished the downstream perturbation from the upstream perturbation. Correspondingly, the cross-stream velocity perturbation map (second row) and the cross-stream tractional force map (fourth row) are not perfectly anti-symmetric about the vertical centerline. In contract, at the highest volume fraction of $\phi = 0.45$ considered, the perturbation maps are far more closer to being symmetric or anti-symmetric about the vertical centerline. This is clearly due to the enhanced interference due to the downstream neighbors that are located in the particle wake.

\section{Force Prediction - Static Case}
This section will consider advanced force models that go beyond the $(\Rey,\phi)$-dependent model by taking into account the precise information on the neighboring particles. In this section, this will however be done only in the static context of flow over a stationary random distribution of particles. The more relevant case of freely moving particles will be considered in the next section. In the static case, the force on the $i^{th}$ particle can be expressed symbolically as
\begin{equation} \label{eq:15}
\bF_i = \bF_{0i}(\Rey_i,\phi_i) + \bF'_i \, ,
\end{equation}
where $\bF_0$ is the standard $(\Rey,\phi)$-dependent force model (eq. given in \eqref{eq:tenneti}) that accurately captures the mean, while $\bF'$ accounts for deviation from the mean by taking into account the precise relative location of the neighbors. In the static case, only the relative location of neighbors matter. Though we consider flow through a uniformly distributed array of stationary particles, in our notation we allow for the macroscale flow, and as a result for the macroscale Reynolds number and average local volume particle fraction to vary from particle to particle. 

In the first subsection, we will consider the pairwise interaction extended point-particle (PIEP) approach, where the superposable wake and the perturbation maps obtained from flow prediction of the previous sections are used to develop the PIEP model. The second subsection will consider an artificial neural network (ANN) approach to directly model the force on a particle based on neighbor information, using the same PR-DNS data used in the development of PIEP model. The third subsection will compare the performance of PIEP and the ANN models using PR-DNS test data that has not been used in the training of the models. 

\subsection{PIEP: Undisturbed Flow Prediction}
\begin{figure}[H]		   	
	\begin{center}
	\includegraphics[width = 5.7in]{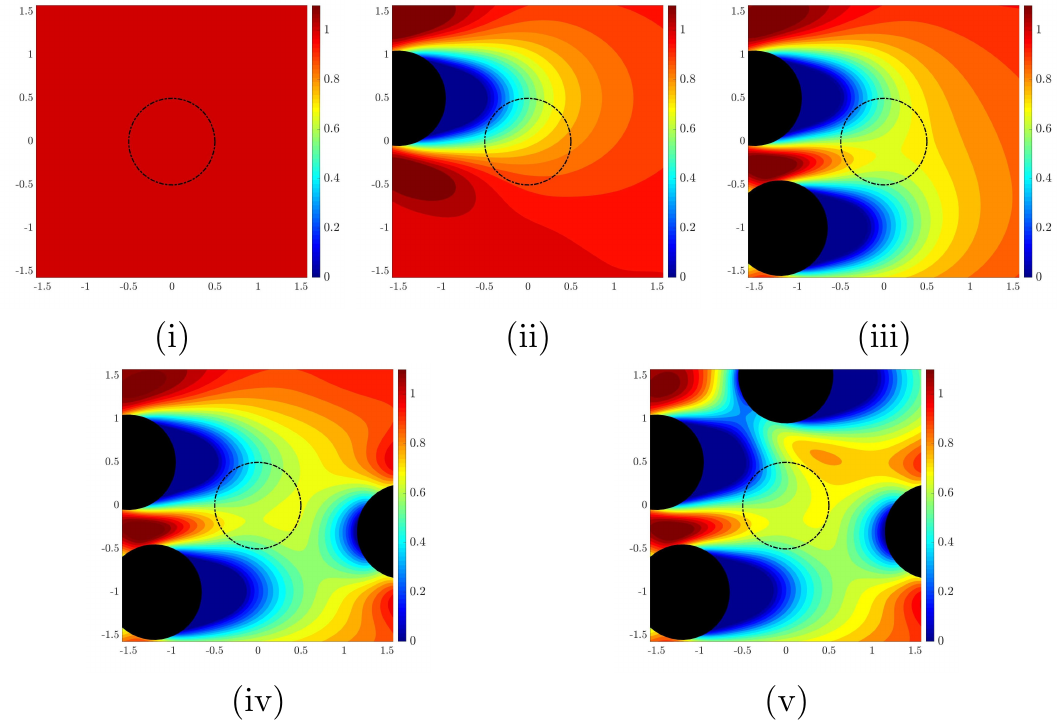}
	\end{center}   
	\caption{Contour plots of streamwise velocity plotted on one $x-y$ plane through the random array of particles, with the $i^{th}$ particle at the center of the frame. Only a portion of the computational domain is shown and depending on where the plane passes, the neighbors appear as circles of different radii. (a) The leading order representation of the undisturbed flow at the $i^{th}$ particle as a uniform flow of unit magnitude and the $i^{th}$ particle is shown as the dash circle. (b) The  undisturbed flow at the $i^{th}$ particle as the sum of the uniform flow plus the perturbation flow induced by the closest neighbor marked as ``1''. (c - e) The  undisturbed flow as the sum of the uniform flow plus the perturbation flows induced by increasing  number of neighbors that are marked in frame. The power of superposable wake in providing accurate representation of the undisturbed flow, by taking into account the influence of neighbors, is clear.}
	\label{fig5}
\end{figure}

We first note that all force models attempt to predict the force on a particle are based on the {\it {undisturbed flow}} of that particle. The undisturbed flow of the $i^{th}$ particle is formally defined as the flow that would exist in the absence of the $i^{th}$ particle, but with all other particles present. Thus in a $N$-particle system, there are $N$ different undisturbed flows - one for each particle, and they differ from each other. As illustrated in Figure \ref{fig5} the undisturbed flow of the $i^{th}$ particle is substantially different from the macroscale flow $\bu_0$ due to the influence of all its neighbors. Figure \ref{fig5} shows the streamwise component of flow around a particle marked ``$i$'' on a plane passing through its center. In this schematic only a small portion of the computational domain centered around the $i^{th}$ particle is shown. Figure \ref{fig5}a shows the macroscale flow (which in our example is a uniform flow of Reynolds number $\Rey_i$ and non-dimensional magnitude unity) with a dash-line denoting the outlines of the $i^{th}$ particle, and all other particles absent from the domain. This is the macroscale-undisturbed flow of the $i^{th}$ particle without accounting for the precise arrangement of its neighbors.  The standard $(\Rey,\phi)$-dependent force model if based on this macroscale-undisturbed flow around the $i^{th}$ particle, and accounts for neighbors only in a statistical sense through volume fraction dependence of  $\bF_{0i}$. 

Here we are interested in including the perturbation effect of each individual neighbor. Figure \ref{fig5}b shows the superposed flow in the region around the location of the $i^{th}$ particle taking into account its closest neighbor marked ``1''. This is an improved estimate of the undisturbed flow of the $i^{th}$ particle that accounts for the perturbation effect of neighbor ``1''. This was computed using \eqref{eq:4} where the summation included only the superposable wake of neighbor ``1'', where the superposable wake is that given in Figure \ref{fig2} for the $\Rey$ and $\phi$ of the neighbor. Figures \ref{fig5}c to \ref{fig5}e present further improved estimates of the undisturbed flow at the $i^{th}$ particle that account for the effects of the nearest two, three and four particles, respectively. Clearly the complexity of the undisturbed flow increases as more neighbors' influence are included in the estimation. In this example, it is clear that the effect of neighbors is to overall decrease the streamwise velocity at the $i^{th}$ particle and also to introduce a small non-zero $y$-component (not shown). This results in a substantially lower drag than the mean and a non-zero lift force. Thus, the complexity of the undisturbed flow due the influence of neighbors, at least partly, can explain the large particle-to-particle force variation observed in \Cref{dragVslift}. As a final point we note that the computation of the undisturbed flow at the $i^{th}$ particle presented in Figure \ref{fig5} requires (i) macroscale information, (ii) relative position {of} each neighbor, and (iii) superposable wake maps presented in Figure \ref{fig2}.  

\subsection{PIEP: Force Prediction from Flow Prediction}
In order to use the perturbation maps of the previous section, we first define three unit vectors: $\bbe_\parallel$, $\bbe_\perp$ and $\bbe_o$, where $\bbe_\parallel$ is the unit vector along the flow direction, which in the static PR-DNS case is taken to be $\bbe_x$. The other two unit vectors are defined as
\begin{equation} \label{eq:16}
\bbe_\perp = \dfrac{(\bx_i -\bx_j) - [(\bx_i - \bx_j) \cdot \bbe_\parallel ]\bbe_\parallel}{|(\bx_i -\bx_j) - [(\bx_i - \bx_j) \cdot \bbe_\parallel ]\bbe_\parallel |} \quad \mathrm{and} \quad 
\bbe_o = \bbe_\parallel \times \bbe_\perp \, ,
\end{equation}
where $\bbe_\perp$ is the in-plane unit-vector normal to $\bbe_\parallel$ in the plane containing $\bbe_\parallel$ and the centers of the $i^{th}$ and the $j^{th}$ particle. The out-of plane direction is $\bbe_o$. 

The next step of the PIEP model is to use the undisturbed flow obtained from flow prediction along with the Generalized Fax\'en relation (or the Maxey-Riley equation) for force prediction. The perturbation force on the $i^{th}$ particle is separated into five contributions as follows
\begin{equation}\label{eq:17}
\bF'_i = \bF'_{un,i} + \bF'_{qs,i} + \bF'_{am,i} + \bF'_{l,i} + \bF'_{c,i} \, .
\end{equation}
The first term on the right hand side is the undisturbed flow and often in the literature it is referred to as the pressure-gradient or stress-divergence ot Archimedes force. Even in the absence of the $i^{th}$ particle this force is realized but the fluid mass that occupies the volume of the $i^{th}$ particle. The other four terms are due to the presence of the $i^{th}$ particle. The second to fourth term are the quasi-steady, added-mass and vorticity-induced lift contributions and they are expressed in terms of the perturbation maps. The last contribution is an additional force correction that will be optimized from the PR-DNS training data with the application of a simple machine learning algorithm (linear regression) and will be described in the following subsection. In the above we have ignored the unsteady viscous (or history) contribution. Following the application of Generalized Fax\'en relation for the undisturbed flow as described in \cite{akiki2017pairwise, akiki2017pairwiseJCP} we obtain the following expression for the undisturbed flow force on the $i^{th}$ particle
\begin{equation}\label{eq:18}
\bF'_{un,i} = \dfrac{\pi}{6} \sum_{j=1, j\ne i}^{N} \left( \mathscr{M}_{f3}(X_{ij},Y_{ij} \,|\, \Rey_j,\phi_j) \, \bbe_\parallel + \mathscr{M}_{f4} (X_{ij},Y_{ij} \,|\, \Rey_j,\phi_j) \, \bbe_\perp \right) \, ,
\end{equation}
where the sum is over the neighbors and can be typically truncated to only ten nearest neighbors. In the above non-dimensional force expression, $\rho u_{0*}^2 d_*^2$ has been chosen as the force scale and $\rho u_{0*}^2$ is the pressure scale. Here the pressure and viscous stress distribution of the undisturbed flow around the $i^{th}$ particle is expressed on a sum of contributions from the perturbation flow of each neighbor. Furthermore, the integration of this force over the $i^{th}$ particle has already been conveniently pre-computed and stored as maps. Note that in the summation the unit vectors are specific to each $(i,j)$ particle pair and as a result $\bF'_{un,i}$ will in general have all three non-zero components.

Similarly the perturbation quasi-steady force on the $i^{th}$ particle can be expressed in terms of the perturbation velocity maps as
\begin{equation}\label{eq:19}
\bF'_{qs,i} = \dfrac{3 \pi}{\Rey} \sum_{j=1, j\ne i}^{N} \left( \mathscr{M}_{f1}(X_{ij},Y_{ij} \,|\, \Rey_j,\phi_j) \, \bbe_\parallel +  \mathscr{M}_{f2} (X_{ij},Y_{ij} \,|\, \Rey_j,\phi_j) \, \bbe_\perp \right) (1 + 0.15 \Rey_i^{0.687}) \, .
\end{equation}
The finite Reynolds number correction of the quasi-steady drag has been taken to be based on the macroscale Reynolds number of the $i^{th}$ particle following the recommendation of \cite{akiki2017pairwise}. The expressions of the added-mass and vorticity-induced lift force are 
\begin{equation}\label{eq:20}
\bF'_{am,i} = \dfrac{\pi \, C_m}{6} \sum_{j=1, j\ne i}^{N} \left( \mathscr{M}_{f3}(X_{ij},Y_{ij} \,|\, \Rey_j,\phi_j) \, \bbe_\parallel + \mathscr{M}_{f4} (X_{ij},Y_{ij} \,|\, \Rey_j,\phi_j) \, \bbe_\perp \right) \, ,
\end{equation}
\begin{equation}\label{eq:21}
\bF'_{l,i} = \dfrac{\pi \, C_L}{6} \sum_{j=1, j\ne i}^{N} \left( \mathscr{M}_{f1}(X_{ij},Y_{ij} \,|\, \Rey_j,\phi_j) \, \bbe_\parallel + \mathscr{M}_{f2} (X_{ij},Y_{ij} \,|\, \Rey_j,\phi_j) \, \bbe_\perp \right) \times \left( \mathscr{M}_{f5}(X_{ij},Y_{ij} \,|\, \Rey_j,\phi_j) \, \bbe_o \right) \, ,
\end{equation}
where $C_m$ is the added-mass coefficient, which is equal to $1/2$ in the low volume fraction limit and can be taken to be a function of volume fraction at finite volume fraction. In the lift force, $C_L$ is the lift coefficient, which can be taken to be $1/2$. The sum of all four perturbation force contributions can be substituted into \eqref{eq:15} to obtain the total force on the $i^{th}$ particle. This process can be repeated to calculate the neighbor-informed force on all the particles within the array and this completes the static implementation of the PIEP force model. 

\subsubsection{PIEP: Quasi-Steady Force Correction Using Regression}
Rigorous test of the PIEP model (to be discussed section \ref{sec:6.3}) shows that the model given in \eqref {eq:17} performs adequately with only the first four contributions described above. The performance can be further improved by incorporating an additional correction whose motivation will be rationalized below. There are significant approximations in both steps of the PIEP model that contribute to errors in its predictions: (i) The undisturbed flow, computed as a superposition of superposable wakes of each neighbor taken one at a time, clearly ignores the effect of nonlinear interactions among the neighbors and (ii) The Maxey-Riley equation (or the generalized Fax\'en form), which expresses the force on the particle in terms of the undisturbed flow, is rigorous only in the limit of an isolated particle in the zero Reynolds number limit.


Within the framework of pairwise interaction, we propose
\begin{equation}\label{eq:18a}
\bF'_{c,i} = \sum_{j=1, j\ne i}^{N} \left( \mathscr{M}_{f6}(X_{ij},Y_{ij} \,|\, \Rey_j,\phi_j) \, \bbe_\parallel + \mathscr{M}_{f7} (X_{ij},Y_{ij} \,|\, \Rey_j,\phi_j) \, \bbe_\perp \right) \, ,
\end{equation}
where the summation is over the neighbors. The streamwise and transverse correction maps, $\mathscr{M}_{f6}$ and $\mathscr{M}_{f7}$, are functions of the separation distance $(X_{ij},Y_{ij})$ between the $i^{th}$ particle and its $j^{th}$ neighbor. These maps are evaluated for each combination of $\Rey$ and $\phi$ and thus are dependent on these two macroscale parameters as well. In \cite{akiki2017pairwiseJCP}, similar corrections were defined based on PR-DNS of two stationary interacting spheres. These corrections are appropriate in the limit of small volume fraction and the results to be presented below are generalization to finite volume fraction.

\begin{figure}[H]		   	
	\begin{center}
		\includegraphics[width = 5.7in]{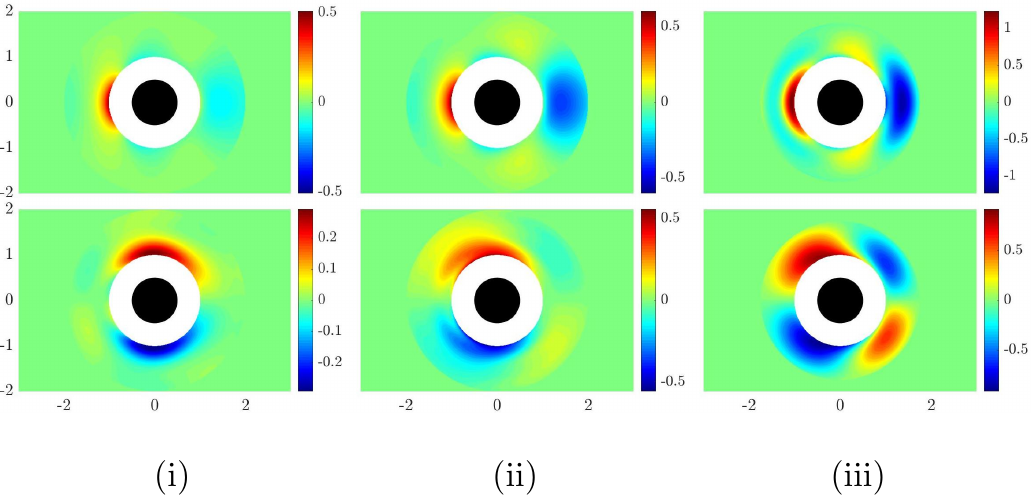}
	\end{center}   
	\caption{Rows corresponding to maps $\mathscr{M}_{f6}$ and $\mathscr{M}_{f7}$. The three columns correspond to the three cases: (i) $\Rey = 39$, $\phi = 0.11$, (ii) $\Rey = 16$, $\phi = 0.21$, and (iii) $\Rey = 21$, $\phi = 0.45$.}
	\label{fig5add}
\end{figure}

The machine learning approach described for the superposable wake is now used for modeling the correction force maps. In particular, $\mathscr{M}_{f7}$ and $\mathscr{M}_{f7}$ are expanded in spherical harmonics (similar to those used for superposable pressure, since there is no divergence-free constraint for the force components). The coefficients of the spherical harmonic expansion are then optimized through linear regression by minimizing the mean square difference 
\begin{equation}\label{eq:17a}
\dfrac{1}{M\, N} \sum_{m =1}^M \sum_{i=1}^N \left[ \bF_{DNS,i} - \left(\bF_{0,i} + \bF'_{un,i} + \bF'_{qs,i} + \bF'_{am,i} + \bF'_{l,i} + \bF'_{c,i} \right) \right]^2 \, .
\end{equation}
The correction maps $\mathscr{M}_{f6}$ and $\mathscr{M}_{f7}$ obtained from linear regression are shown in the two rows of Figure \ref{fig5add}, where the three columns correspond to the three cases shown in Figure \ref{fig4}. We emphasize the approximate nature of the correction due to the assumption of pairwise superposition and furthermore the model is based on linear regression that uses limited amount of PR-DNS data. 

\subsection{Artificial Neural Network (ANN) for Direct Force Prediction}
The work of He \& Tafti \cite{he_tafti} is of particular relevance. They constructed a multi-layer feed-forward network to predict drag variation on stationary spherical particles. Here we essentially follow their approach to develop an ANN model of the particle force and compare its prediction against those from the PIEP model. In the present study, the training of the ANN uses the PR-DNS data of the eight cases listed in Table 1 that are identical to those used in the training of the superposable wake in the PIEP model. For each particle within the array, the inputs to the ANN are the macroscale $\Rey$ and $\phi$ of the particle and its neighbors' locations. The output of the ANN is the drag force on the particle.   

\begin{figure}[H]		   	
	\begin{center}
		\includegraphics[width = 3in]{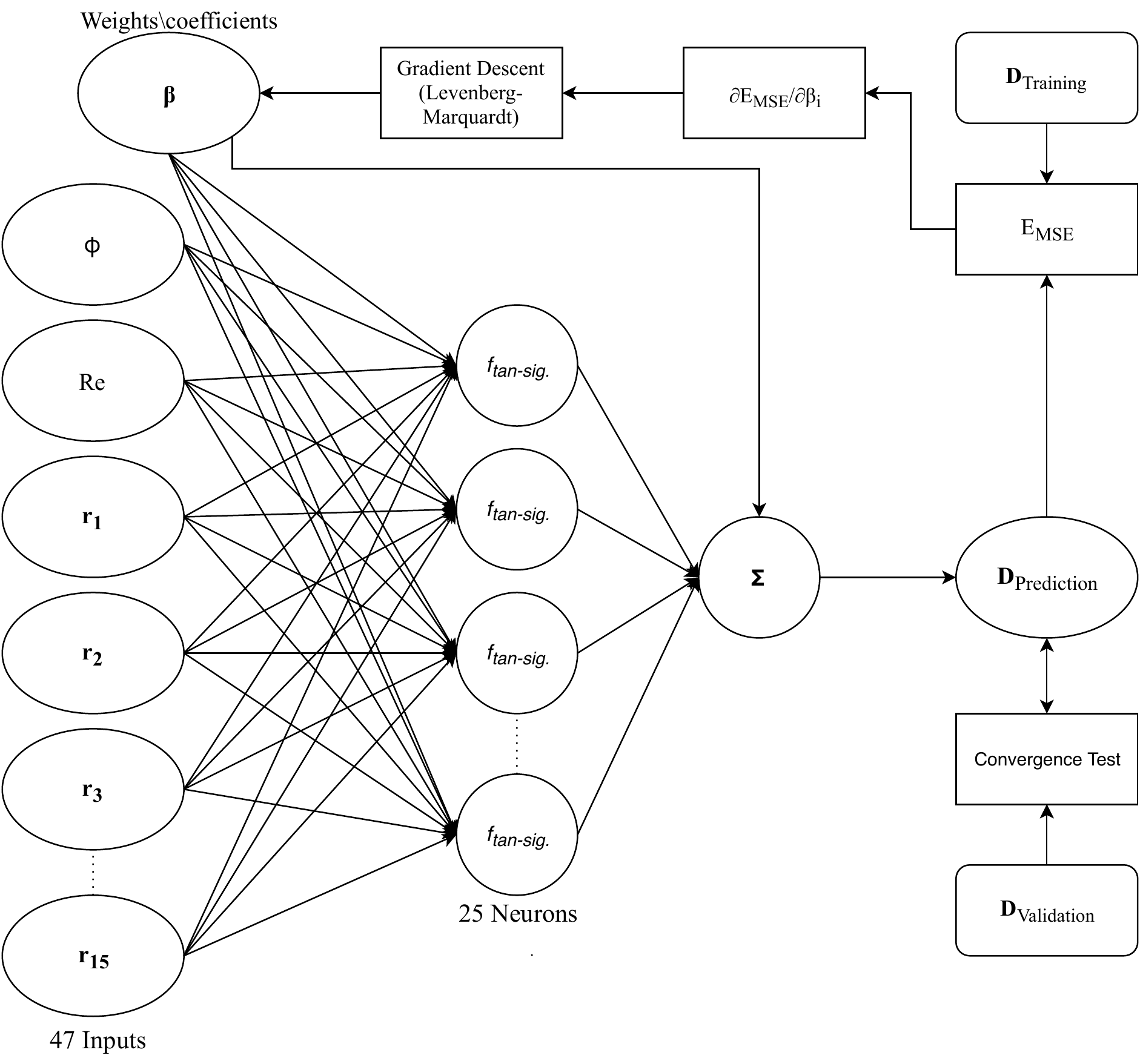}
	\end{center}   
	\caption{Structure of the multi-layer feed-forward network used here which is the similar to that employed in \cite{he_tafti}.}
	\label{fig_ann_structure}
\end{figure}

The structure of the network is presented in Figure \ref{fig_ann_structure}. We define the difference between the actual drag on a particle and the corresponding ANN prediction as the drag error (the drag is the streamwise component of force). The cost function is defined as the root mean square drag error over all the training sample, normalized by the mean of the actual drag from the PR-DNS. We follow the recommendation of \cite{he_tafti}, who observed inclusion of 15 neighbors to provide the optimal prediction. Thus, for each particle the input vector consists of 47 features ($\Rey$, $\phi$ and 15 relative position). The hidden layer of the ANN contains 25 neurons with the hyperbolic tangent sigmoid activation function. The hidden layer is followed by an output layer with a single scalar value of particle drag as the output. The Levenberg-Marquardt back-propagation method is used to train the ANN and the details are identical to those recommended in \cite{he_tafti}.

Since the data set used in \cite{he_tafti} and here are not the same, the resulting ANNs will be biased towards the different $\Rey$, $\phi$, and particle configurations of the PR-DNS data that were used in the respective training process. Nevertheless, by following their methodology, we trained an ANN using PR-DNS of flow through a stationary, monodisperse array of particles. Rigorous testing of the ANN drag model along with the testing of the PIEP model will be discussed in the next subsection. For this testing we withheld a single PR-DNS realization for each $\Rey$ and $\phi$ combination. The remaining PR-DNS realizations consisted of a large number of particles, which were used in the following fashion: 70\% were used for training of the ANN, 15\% for validation, and 15\% for pre-testing. Both the training and validation data are used by the Matlab ANN code in developing the ANN model (see \ref{fig_ann_structure}), where the training data is used for optimizing the coefficients of the neural network nodes, while the validation data is used for evaluating convergence. The pre-testing data is used only after the training of the ANN. There is an important distinction between the testing and pre-testing data:  the pre-testing data comes from the same realizations as the training and validation data (although the data itself is different), and the testing data comes from completely separate realizations. In this manner, the pre-testing data is not a true ``blind" test since data from the same realization were used in training the model. This is important since the ANN performs significantly better with the pre-testing data vs. the testing data. The results to be presented are reasonably invariant to changes in the various activation functions, hyper parameters and network architectures.

\begin{figure}[H]		   	
	\begin{center}
		\includegraphics[width = 6.7in]{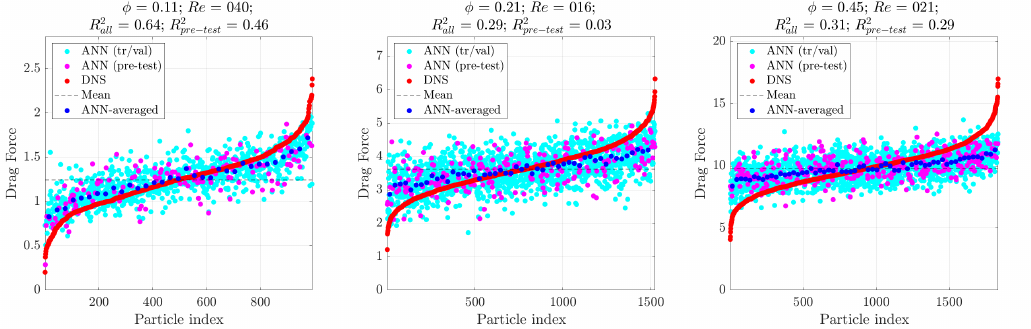}
	\end{center}   
	\caption{All but one PR-DNS realizations are randomly partitioned into training/validation and pre-testing data. The particle index has been sorted form low- to high-drag particles. The blue dots represent a moving average of both training and testing data where the averaging window is 30 particle indices.}
	\label{fig_ann_train}
\end{figure}

The drag prediction by the ANN model for both the training/validation data (cyan symbols) and pre-test data (pink symbols) are reported in Figure \ref{fig_ann_train} along with the actual drag measured in the PR-DNS (red symbols). Also reported in the figure are the corresponding $R^2$ values computed as
\be
R^2 = 1- \cfrac{\sum_{j=1}^{N_{samples}} \, |F_{PR-DNS}(j)-F_{model}(j)|^{2}}{\sum_{j=1}^{N_{samples}} \,|F_{PR-DNS}(j)-\langle F_{PR-DNS} \rangle|^{2}}  \label{eq:r2} \, .
\ee
The $R^2$ values reported in the figure are for all the particles, as well as only for the pre-test data that was not used in the training process. Similar to observations by \cite{he_tafti}, the overall trend of the ANN prediction matches that of the PR-DNS data. That is, the ANN correctly identifies the particles whose drag is lower and higher than the mean. However this qualitative agreement with the PR-DNS data is dominantly restricted to the training data. Even for the training data, quantitative difference persists. For example, the distribution of ANN prediction (cyan symbols) away from the average value is not as large as that exhibited by the PR-DNS drag. This can be better seen in the difference between the blue and the red curves, where the blue symbols correspond to local average drag over 30 particles. ANN loses its predictive capability when applied to the pre-test data, which was not used in the training/validation process, as reflected by the much lower $R^2$ values. This indicates that the amount of PR-DNS data used for training is not sufficient and as a result the ANN is over-fitting the training data. This is not a problem with the ANN itself, but it indicates that this issue can be remedied only with lot more PR-DNS data. This is what we earlier mentioned as the challenge to direct force prediction, without the intermediate step of flow prediction.

\subsection{Static Test of PIEP and ANN Models} \label{sec:6.3}
To compare the performance of the PIEP model to the ANN, both models are applied to the PR-DNS realization that was not used in the training of either model. In this manner, the models to be tested are relatively unbiased towards the test DNS realization. The resulting $R^2$ values are given in Table \ref{tab:r2}. The PIEP model results for the force in the streamwise and transverse direction are given here along with the ANN's force prediction. It is important to emphasize the difference between pre-testing and testing to be reported in this section. In case of pre-testing, the drag data of particles used for pre-testing came from the same PR-DNS simulations that were used for training. I.e., though the training and pre-testing data were separate they came from the same PR-DNS simulations. In contrast, here we will attempt to predict the drag from an entirely new PR-DNS realization whose data has not been used in any way before.

\begin{table} [H]
	\begin{center}
		\begin{tabular}{ 
				|p{1cm}|p{1cm}|p{1cm}||p{2.2cm}|p{2.2cm}|p{2.2cm}||p{2.2cm} }
			\hline
			Case & $\phi$ & $\Rey$ & $F_{x,ANN}$ & $F_{x,PIEP}$ & $F_{y,ANN}$ & $F_{y,PIEP}$  \\
			\hline	
			\hline	
			1 & 0.11  & 39  & -0.30 & 0.65 & 0.13 & 0.75 \\
			2 & 0.11  & 70  & -0.21 & 0.59 & 0.12 & 0.71  \\
			3 & 0.11  & 173 & -0.30 & 0.53 & 0.03 & 0.64  \\
			\hline
			4 & 0.21  & 16  & -0.26 & 0.75 & 0.03 & 0.75  \\
			5 & 0.21  & 86  & -0.12 & 0.68 & 0.09 & 0.65  \\
			\hline
			6 & 0.45  & 3   & -0.28 & 0.55 & -0.08 & 0.57 \\
			7 & 0.45  & 21  & -0.24 & 0.59 & -0.03 & 0.58 \\
			8 & 0.45  & 115 & -0.17 & 0.64 & 0.04 & 0.57  \\
			\hline
		\end{tabular}
		\caption{Resulting $R^2$ values for the ANN and PIEP model for a realization not used in model development}
		\label{tab:r2}
	\end{center}
\end{table}	

\begin{figure}[H]		   	
	\begin{center}
		\includegraphics[width = 6.7in]{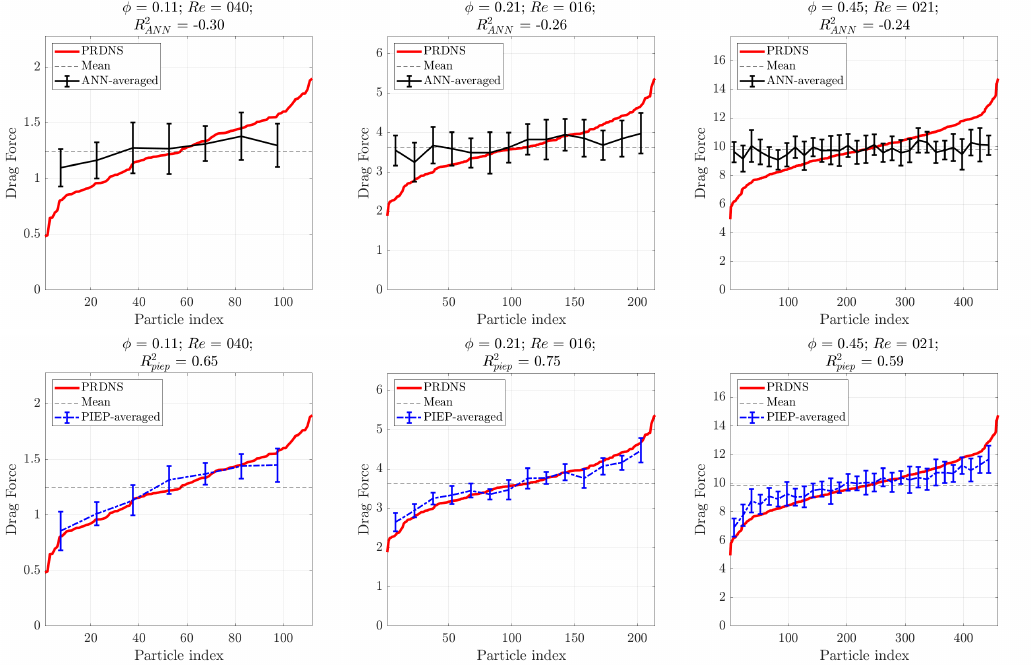}
	\end{center}   
	\caption{Results of the neural network and the PIEP model when applied to a realization not used for developing either model. The particle index has been sorted form low- to high-drag particles. The averaging window is 15 particles indices due to the smaller number of particles. {The vertical bars indicate the standard deviation in each direction of the predictions taken over the averaging window.}}
	\label{fig_ann_test}
\end{figure}

When the models are applied to the test realization, the difference between the ANN and the PIEP model becomes even more evident. The $R^2$ values for the ANN{'s drag} are negative for all $\Rey$ and $\phi$ combinations. Once again, this indicates that the ANN over-fits the training data. Since the training data and the pre-test data were from the same realizations, some information was shared between them. In contrast, when a completely independent realization is used for testing, the ANN predictions become worse than even the mean-drag model, as reflected by the negative $R^2$ value. This over-fitting behavior is further demonstrated in Figure \ref{fig_ann_test}. The ANN is not able to accurately identify the spheres with substantially above and below average drag. On the other hand, the PIEP model retains its predictive capability even when applied to the test realization that was not used in the training process. The PIEP model is able to capture approximately 53\% to 75\% of the variations in particle forces depending on $\Rey$ and $\phi$. The ANN prediction remains close to the average indicating its inability to capture positive and negative deviations from the average, while the running  average of the PIEP model was able to capture the DNS trend quite well. Again this should not be interpreted as a fundamental weakness of ANN's ability to arrive at an accurate model. This merely indicates the need for a much larger amount of PR-DNS data if we want to pursue direct force prediction.

The above presented PIEP model results are with all five contributions given on the right hand side of \eqref{eq:17}, i.e., including the correction term that was modeled using linear regression to account for the higher-order effects. Without the correction term, the performance of the PIEP model somewhat decreases, but nevertheless the performance is still far better than the ANN model. We have also tried a limiting case of PIEP prediction, where in \eqref{eq:17} only the correction term $\bF'_{c,i}$ was retained (i.e., without the unsteady flow, quasi-steady, added-mass and lift contributions obtained from flow prediction). In this limiting case, the streamwise and transverse correction maps, $\mathscr{M}_{f6}$ and $\mathscr{M}_{f7}$, defined in \eqref{eq:18a} were obtained by approximating $\bF_{DNS,i} - \bF_{0,i}$. We observe the performance of this model to be poor. This is understandable, since such a limiting PIEP model without flow prediction becomes a direct approach and is similar to the ANN model. The prediction can be expected to be worse than that of ANN, since the model suffers from both the limited size of the PR-DNS data and the pairwise superposition approximation.

\section{Dynamic Case - Sedimenting Particles}
In this section we extend the {PIEP} force prediction framework to an EL simulation of a random distribution of particles that freely move in response to the hydrodynamic force acting on them. The setup we consider is an EL simulation of particle-laden flow in a triply periodic box where the particles settle down in the vertical direction in response to the gravitational and hydrodynamic force acting on them. Before we consider the simulation results, the neighbor-dependent PIEP force model for this dynamic case will be discussed.

The multiphase flow is now time dependent and in the EL simulation, the macroscale velocity field $\bu_0(\bx,t)$ is obtained by solving the filtered continuous phase governing equations. Our quest is to accurately model the hydrodynamic force on all the particles within this flow by accounting not only for the macroscale flow, but also for the microscale perturbation flow induced by neighbors. At any given time the state of the particle distribution is characterized by the particle position $\bx_i(t)$, velocity $\bv_i(t)$ and acceleration $d\bv_i/dt$, for $i = 1, 2, \cdots N$. The relative velocity and the relative acceleration between the fluid and the particle are then 
\begin{equation}\label{eq:22}
\bu_{r,i} (t) = \overline{(\bu_0)}^S_i - \bv_i(t) \quad \mathrm{and} \quad \dfrac{d \bu_{r,i}}{dt} = \overline{ \left( \frac{D\bu_0}{Dt} \right) }^V_i - \dfrac{dv_i}{dt} \, ,
\end{equation}  
where $D/Dt$ and $d/dt$ denote total derivatives following the fluid and the particle, respectively. In the above, $\overline{(\bu_0)}^S_i$ indicates surface average of the macroscale fluid velocity over the surface of the $i^{th}$ particle and $\overline{ \left( {D\bu_0}/{Dt} \right) }^V_i$ denotes the volume average of the fluid acceleration at the $i^{th}$ particle. Thus, in the definition of relative velocity, the undisturbed fluid velocity of the $i^{th}$ particle has been taken to be the surface average. This definition is analytically precise in the zero Reynolds number limit from the Fax\'en's theorem, and its use at finite Reynolds number is an approximation. The particle Reynolds number $\Rey_i$ is obtained based on this relative velocity. Also, the distribution of particles within the computational domain will vary over time and let $\phi(\bx,t)$ be the macroscale particle volume fraction field obtained by filtering the particle location \cite{cap_desjardin, bala_liu}. With this the macroscale particle volume fraction around the $i^{th}$ particle can be obtained as $\phi_i(t) = \phi(\bx_i,t)$. 

\subsection{Flow Prediction at the Microscale}
\begin{figure}[H]		   	
	\begin{center}
		\includegraphics[width = 5in]{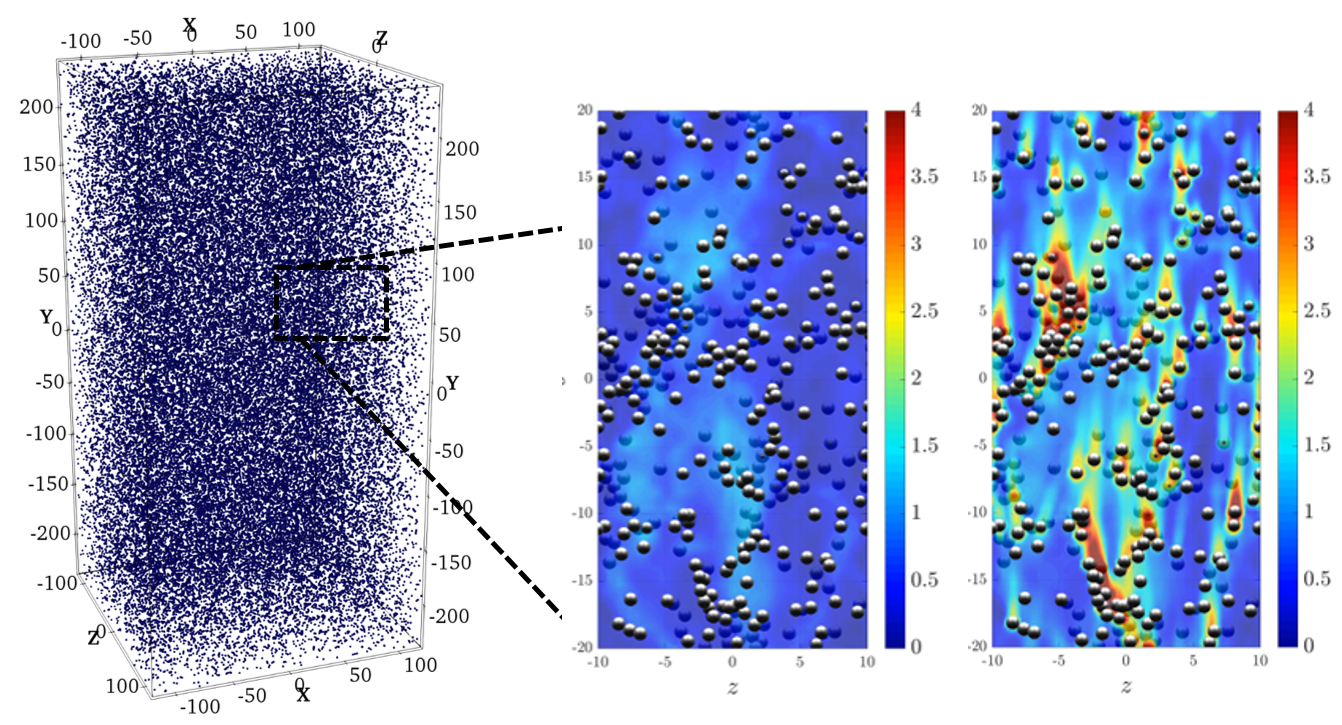}
	\end{center}   
	\caption{Results of an EL simulation of sedimentation of 168972 particles of unit diameter in a triply periodic box of size $192 \times 96 \times 96$, which corresponds to a volume fraction of 5\%. (a) Schematic of the problem set up and the particle distribution at one time instance. (b) Contour plot of vertical velocity computed in the EL simulation plotted on a vertical plane in a small region of the computational domain. Also shown are particles that are close to this plane. The particles are given a finite size, even though the EL simulations considered the particles to be points and their size mattered only in the soft-sphere collision algorithm. Due to two-way coupling the flow varies on a scale much larger than the particle size. (c) Contour plot of total vertical velocity computed as the sum of macroscale velocity shown in frame (b) plus the perturbation flow due to each particle obtained in terms of its superposable wake.}
	\label{fig9}
\end{figure}

Figure \ref{fig9}a shows the configuration of {an} EL simulation of sedimenting particles in a triply periodic box of size $192 \times 96 \times 96$ along the vertical and the two horizontal directions. The computational domain consisting of 168972 particles of unit diameter, which corresponds to an average volume fraction of 5\%. The gravitational velocity $U_{g*}=\sqrt{(\rho_{p*}/\rho_{f*}-1)d_*g_*}$ is chosen as the reference velocity scale and the Reynolds number based on it is $\Rey=178.46$ (particle diameter is the length scale). The EL simulations employed a Gaussian filter of size $\delta=3$. The filtered governing equations of the fluid phase are solved using spectral element methodology using the code Nek5000 \cite{zwick_bala}. Cuboidal spectral elements of size $12 \times 12 \times 12$ are employed in the computations with each element discretized by $17 \times 17 \times 17$ Gauss-Lobatto Legendre grid points. The number of spectral elements used in the vertical direction are 16, while the other two directions employed 8 spectral elements. Thus, the EL simulations employed a spectral element grid of over 5 millions of grid points. Figure \ref{fig9}a shows the computational domain with the position of all the particles marked at one instant in time.

Figure \ref{fig9}b shows a zoom-up of a small region marked by a red box in Figure \ref{fig9}a. Only a  vertical plane is shown in which contours of vertical component of the macroscale fluid velocity $\bu_0(\bx,t)$ computed in the EL simulation is shown (vertical velocity of fluid is defined such that its mean averaged over the entire box is zero and is positive directed downward). Also plotted in this figure are the particles that intersect this plane. As far as the fluid flow is concerned, the particles are point particles, and thus the macroscale flow is solved over the entire computational domain. For visual effect the particles are given a finite size in Figure \ref{fig9}b. It should be noted that the EL simulations employ a soft-sphere collision algorithm where the particles are assumed to be of finite size. It can be seen that in the present EL simulation, the macroscale velocity varies over the computational domain and this variation is due to local clustering of particles and their feedback force on the fluid.  However, since the Gaussian filter size is much larger than the particle diameter, the macroscale flow in Figure \ref{fig9}b varies on a scale broader than the particle diameter, and in particular microscale features such as boundary layers and wakes around the particles cannot be observed in the flow. The vertical plane shown in the zoom-up of Figure \ref{fig9}b employed a non-uniform Gauss-Lobatto grid of $26 \times 52$ grid points. 

The superposable wakes developed in the previous sections can now be used for flow prediction of the microscale details, such as boundary layers and wakes around the particles. The perturbation flow due to the $i^{th}$ particle is characterized by its Reynolds number $\Rey_i$ and the local particle volume fraction $\phi_i$. Within the framework of pairwise interaction, the perturbation flow of each particle is independent of the precise spatial distribution of all other particles and can be taken to be the superposable wake presented in Figure 3. A superposition of superposable wakes due to all the particles within the system will correspond to the total microscale perturbation flow. This microscale flow can then be added to the macroscale flow computed in the EL simulation, as shown in \eqref{eq:2}. Two adjustments must however be made: (i) The axisymmetric superposable wake of each particle will now be oriented in the direction of instantaneous relative velocity $\bu_r$ of that particle. In the static example of section 6, the relative velocity of all the particles were taken to be the same, oriented along the $x$-axis. This will not be the case in an EL simulation of freely moving particles. A coordinate transformation from the wake-aligned coordinates in which $\bu_{sw}$ is defined to the global coordinate system is needed to construct the sub-grid microscale flow of an EL simulation. (ii) Due to the lower volume fraction of the system, the force correction $\bF'_{c,i}$ is taken to be that given by two-particle DNS as given in \cite{akiki2017pairwiseJCP}. (iii) The superposable wake presented in Figure 3 was normalized by relative velocity magnitude. Therefore, in the superposition process, the wake of each particle must be re-scaled to the velocity scale of the EL simulation. With these adjustments the superposed velocity and pressure can be expressed as
\begin{equation}\label{eq:23}
\bu(\bx,t) = \bu_0(\bx,t) + \sum_{j=1}^N |\bu_{r,j}| \, \bQ_j \, \bu_{sw}(r_j,\theta_j \, | \, \Rey_j , \phi_j) \, ,
\end{equation}
\begin{equation}\label{eq:24}
p(\bx,t) = p_0(\bx,t) + \sum_{j=1}^N |\bu_{r,j}|^2 \, p_{sw}(r_j,\theta_j \, | \, \Rey_j , \phi_j) \, ,
\end{equation}
where $r_j = |\bx - \bx_j|$ is the distance from the center of the $j^{th}$ particle and the angle $\theta_j = \cos^{-1}(((\bx-\bx_j) \cdot \bu_r)/(r|\bu_r|))$. The rotation matrix $\bQ_j$ transforms the velocity from the wake-attached coordinate to the EL coordinates of the computational box shown in Figure \ref{fig9}a. Also, the velocity perturbations have been scaled by the scaling factor $|\bu_{r,j}|$ and the pressure perturbations have been scaled accordingly.

Figure \ref{fig9}c  shows the zoomed-up small region marked by the red box in Figure \ref{fig9}a, but with contours of the vertical component of the total velocity on the vertical plane. The difference between the macroscale fluid velocity $\bu_0(\bx,t)$ shown in Figure \ref{fig9}b and the total velocity $\bu(\bx,t)$ shown in Figure \ref{fig9}c is clear. The difference is due to the perturbation flow induced by all the particles that are in the neighborhood of the vertical plane. In order to resolve the microscale features such as boundary layers and wakes around the particles, the vertical plane shown in Figure \ref{fig9}c employs a grid that is finer than that used in the EL simulation. 

If the level of resolution used in the EL simulation were to be comparable to that used in the PR-DNS simulations (i.e, ~30 grid points per diameter), such a fully-resolved PR-DNS simulation of the sedimentation problem over the entire domain shown in Figure \ref{fig9}a would require 47.8 billion grid points. Only with such resolution the force on the particles can be calculated from first principle through surface integration of pressure and shear stress. This will then allow accurate evaluation of the particle motion and the induced flow and so on. However, such an highly resolved PR-DNS simulation is prohibitively expensive. The PIEP force prediction model, by including the effect of the microscale perturbation flow due to the particles, in essence, calculates the force on the particles as if the microscale flow around the particles has been computed at a much higher resolution than that used in the underlying EL simulation. It must be stressed that by computing the perturbation maps shown in Figure 4, we avoid the need to explicitly compute the flow $\bu$ on a very fine grid. The steps involved in the neighbor-dependent PIEP force prediction will be described in the following subsection.

\subsection{PIEP Force Prediction}
The standard ($\Rey,\phi$)-dependent force on the $i^{th}$ particle is obtained as $\bF_0(\Rey_i,\phi_i)$. In the dynamic case, the undisturbed flow, quasi-steady, added-mass, and vorticity-induced lift contributions of force due to perturbation influence of neighbors are obtained as follows. First, to account for the axisymmetric perturbation flow of the $j^{th}$ neighbor, we define its instantaneous wake to be oriented along the direction of relative velocity and the parallel unit vector is now defined as
\begin{equation}\label{eq:25}
\bbe_\parallel = \dfrac{\bu_{r,j}}{|\bu_{r,j}|} \, . 
\end{equation}
The other two unit vectors' definitions remain the same as given in \eqref{eq:16}. Note that the orientation of the three orthogonal unit vectors now depend not only on the particle pair, but also on time due to particle motion. In addition, the macroscale Reynolds number and volume fraction of each particle varies over time, as the particles move and constantly rearrange within the flow. In a time evolving EL simulation, the undisturbed flow force on the $i^{th}$ particles can be expressed as 
\begin{equation}\label{eq:26}
\bF_{un,i} = \dfrac{\pi}{6}\left[ - \overline{(\nabla p_0)}^V_i + \dfrac{1}{\Rey} \overline{(\nabla^2 \bu_0)}^V_i \right] + \dfrac{\pi}{6} \sum_{j=1, j\ne i}^{N} |\bu_{r,j}|^2 \left( \mathscr{M}_{f3} \, \bbe_\parallel + \mathscr{M}_{f4} \, \bbe_\perp \right) \, .
\end{equation}
Note $\overline{(\nabla p_0)}^V_i$ indicates volume average of the macroscale pressure gradient over the volume of the $i^{th}$ particle and a similar definition applies for the viscous stress as well. Since the particles are typically much smaller than the EL filter width, the volume average can be replaced by the function evaluated at the center of the $i^{th}$ particle. The first term on the right corresponds to the unsteady flow force due to the macroscale flow, where $\Rey$ is the Reynolds number corresponding to the length and velocity scales chosen for non-dimensionalization. The second term corresponds to contribution from the perturbation flow of neighbors. In the second term, the $(X_{ij},Y_{ij} \,|\, \Rey_j,\phi_j)$-dependence of the perturbation maps has been suppressed. Apart from the time dependence of these parameters and the unit vectors, and the scaling by the square of relative velocity magnitude, the form of the perturbation contribution is the same as in the static case given in \eqref{eq:18}. It must be cautioned that the above expression assumes all the particles to be of the same size and the diameter of the particle to be the length scale.

In the work of \cite{akiki2017pairwiseJCP} the quasi-steady force on the $i^{th}$ particle has been approximated as the sum of the macroscale contribution and the perturbation contribution as
\begin{equation}\label{eq:27}
\bF_{qs,i} = \bF_0(\Rey_i,\phi_i) + \dfrac{3 \pi}{\Rey} \sum_{j=1, j\ne i}^{N} |\bu_{r,j}| \left( \mathscr{M}_{f1} \, \bbe_\parallel + \mathscr{M}_{f3} \, \bbe_\perp \right) (1 + 0.15 \Rey_i^{0.687}) \, .
\end{equation}
Again, the Reynolds number in the denominator is simply based on the reference velocity and the particle diameter. As before, the finite Reynolds number correction of the quasi-steady drag is based on the macroscale Reynolds number of the $i^{th}$ particle. The expressions of the added-mass and vorticity-induced lift force are 
\begin{equation}\label{eq:28}
\bF_{am,i} = \dfrac{\pi \, C_m}{6} \left[ \overline{ \left( \dfrac{D\bu_0}{Dt} \right) }^V_i - \dfrac{d\bv_i}{dt} + \sum_{j=1, j\ne i}^{N} |\bu_{r,j}|^2  \left( \mathscr{M}_{f3} \, \bbe_\parallel + \mathscr{M}_{f4}  \, \bbe_\perp \right) \right] \, ,
\end{equation}
\begin{equation}\label{eq:29}
\bF_{l,i} = \dfrac{\pi \, C_L}{6} \left[ \bu_{r,i} \times \left( \overline{(\bomega_0)}^V_i + 2 \bOmega_i \right) + \sum_{j=1, j\ne i}^{N} |\bu_{r,j}|^2 \left( \mathscr{M}_{f1} \, \bbe_\parallel + \mathscr{M}_{f3} \, \bbe_\perp \right) \times \left( \mathscr{M}_{f5} \, \bbe_o \right) \right] \, .
\end{equation}
The expression for the added-mass force now includes contributions both form the acceleration of the macroscale flow at the location of the $i^{th}$ particle and from the acceleration of the particle. Similarly, the first term on the right hand side of the lift force expression corresponds to the macroscale contribution, where $\overline{(\bomega_0)}^V_i$ is the vorticity of the macroscale flow computed in the EL simulation averaged over the volume of the $i^{th}$ particle and $\bOmega_i$ is the angular velocity of the $i^{th}$ particle.

In the EL simulation, the position and velocity of all the particles are advanced in time by solving the equations of motion of the particles (see \cite{akiki2017pairwiseJCP} for details). The force on each particle consists of hydrodynamic contribution due to the surrounding flow and collisional contribution that is calculated using the soft-sphere algorithm. The hydrodynamic force on a particle is computed as the sum of macroscale contribution and the effect of perturbation flow due to all the neighbors as given above. 

The accuracy of the PIEP force model has been tested in \cite{akiki2017pairwiseJCP} in the context of drafting-kissing-tumbling (DKT) problem of two falling and interacting particles, in the context of a falling square lattice of {five} particles, and also in the context of falling 80 particles initially randomly distributed within a small cubic volume. In all these cases, PR-DNS was performed to serve as reference against which the results of EL simulations with PIEP force model was compared. Due to the small number of particles in these simulations, the background macroscale flow $\bu_0$ was chosen to be quiescent. These examples demonstrated the substantial improvement in the accuracy of particle motion with the inclusion of fluid-mediated particle-particle interaction (i.e., the influence of perturbation flow due to neighbors). 

In the present problem of sedimentation of {168972} particles, a PR-DNS is not possible to serve as reference. Instead, we compare the solution of EL simulations with and without the PIEP model. To compare the two simulations, the metric we choose is the rate of inter-particle collisions that occur in the system. This is particularly a sensitive metric, since, the rate of collision is greatly altered by each particle being informed of the perturbation effect of its neighbor. This influence is most clear in the extreme two-particle case of the DKT problem. Considering the problem of two spheres falling through still fluid, in the inline configuration of one particle behind the other, in an EL simulation, if each particle's drag is calculated without the influence of the other particle, then both particles will continue to fall at the same rate and there will be no collision. However, with the PIEP model, if the upstream particle's wake effect is included in reducing the drag on the downstream particle, it will fall faster than the upstream particle, and soon will collide with it. 

\begin{figure}[H]		   	
	\begin{center}
		\includegraphics[width = 5in]{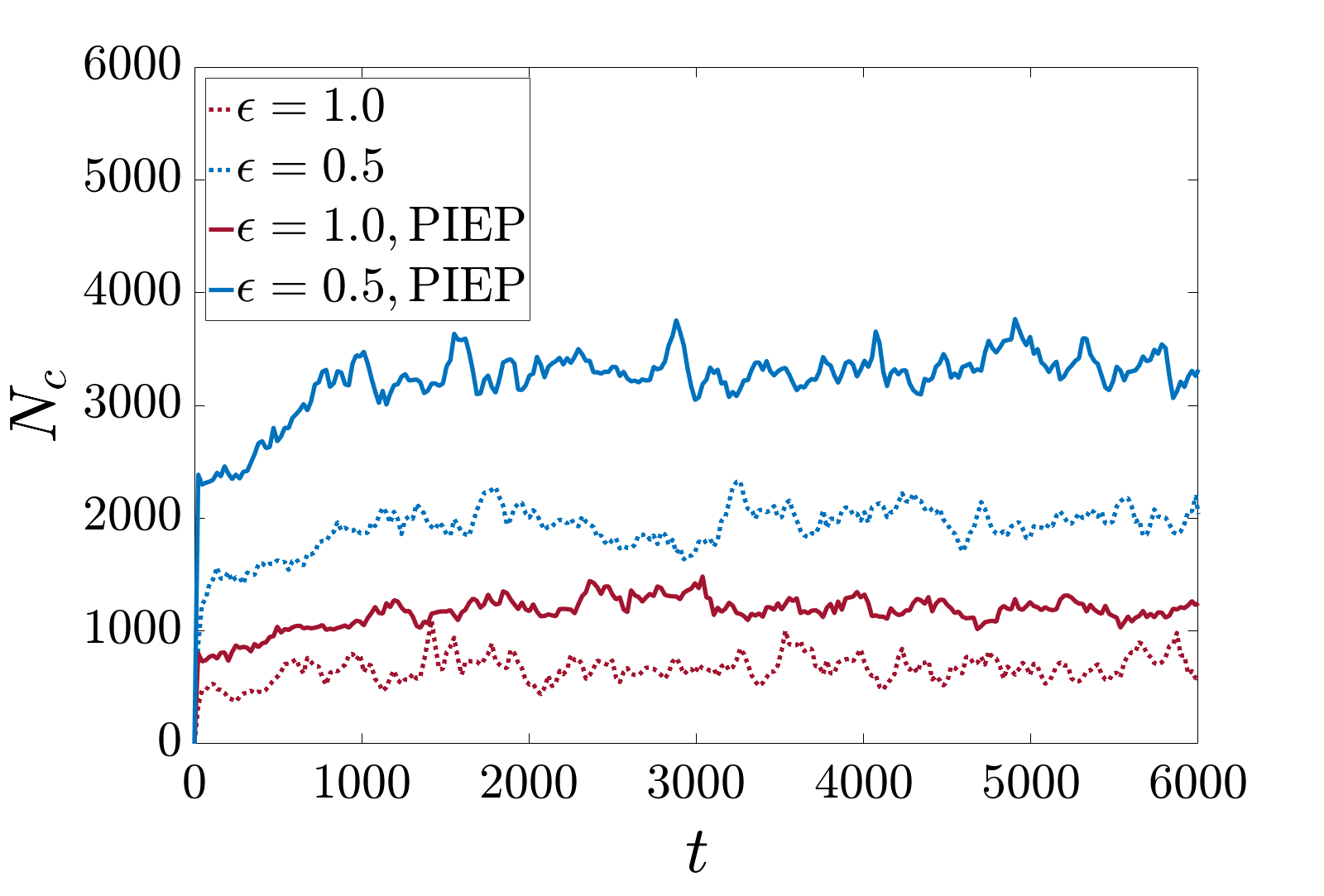}
	\end{center}   
	\caption{Number of collisions per unit time plotted as a function simulation time measured in the EL simulation with and without the PIEP model. Two different coefficient of restitution $\epsilon =1.0$ and $0.5$ are considered.}
	\label{fig10}
\end{figure}

This simple physics continues to be active in a large system of sedimenting particles. Thus, with the PIEP model, particles that are in the wake of upstream particles tend to fall faster and collide with their upstream neighbor, far more frequently than without the PIEP model. This difference is shown in Figure \ref{fig10}, where rate of inter-particle collision is plotted as a function of time for EL simulations with and without PIEP. In the simulation without the PIEP model, the hydrodynamic force on the $i^{th}$ particle includes only the macroscale contribution represented by the first term on the right hand side of equations \eqref{eq:26} to \eqref{eq:29}. In the soft-sphere collision model (see \cite{akiki2017pairwiseJCP}), two different coefficients of restitutions were considered (i.e., $\epsilon =1.0$ and $0.5$). With microscale fluid-mediated inter-particle interactions being evaluated with the PIEP model, the number of particle collisions per time step is nearly doubled. This result is in qualitative agreement with the aforementioned DKT case, in which nearby particles tend to approach each other resulting in collision. On the other hand, coefficient of restitution also plays an important role in particle interaction. Specifically, lower restitution means higher energy damping, which contributes to clustering and thus increases collision rate.

\section{PIEP Modeling of Torque}
The flow prediction using superposable wakes, described in sections 3 and 4, is quite powerful and can be used for modeling other quantities as well. Following the steps used in force prediction, the superposable wakes will be used for the prediction of perturbation torque exerted on a particle due to the influence of its neighbors. Again we consider the perturbing particle to be the $j^{th}$ particle whose axisymmetric wake was shown in Figure 3. The first step of PIEP torque modeling is to place the $i^{th}$ particle at a distance $(X_{ij},Y_{ij})$ away from the center of the $j^{th}$ particle along the $x$ and $y$ directions and define the following two additional {\it {perturbation maps}}:
\begin{equation}\label{eq:30}
\mathscr{M}_{t1}(X_{ij},Y_{ij} \,|\, \Rey,\phi) = \dfrac{6}{\pi} \int_V (\nabla \times \bu_{sw} ) \cdot \bbe_z \, dV \,,
\end{equation}
\begin{equation}\label{eq:31}
\mathscr{M}_{t2}(X_{ij},Y_{ij} \,|\, \Rey,\phi) = \dfrac{1}{\pi} \int_S  \left( \left[ \bn \cdot \dfrac{(\nabla \bu_{sw} + \nabla \bu_{sw}^T)}{2}  \right] \times \bn \right) \cdot \bbe_z \, dS \,.
\end{equation}
where the integrals are over the volume and the surface of the $i^{th}$ particle and $\bn$ is the unit normal vector on the surface of the particle. From the above definitions it is clear that $\mathscr{M}_{t1}$ represents the macroscale vorticity averaged over the volume of the particle and $\mathscr{M}_{t2}$ represents the net  torque exerted by the macroscale flow on a unit sphere of fluid centered about the point $\bx_i$, which is the center of $i^{th}$ particle. Both these torques are oriented in the $z$-direction (i.e., they are perpendicular to the plane formed by the two particle centers and the ambient flow direction).

Just like the five force perturbation maps given in \eqref{eq:10} to \eqref{eq:14}, the above two torque maps are also function of the separation between the center of the $i^{th}$ particle from the center of the disturbing $j^{th}$ particle. They are also functions of $\Rey$ and $\phi$ through the dependence on the superposable wake which was used in their computation. The seven perturbation maps for the different combinations of $\Rey$ and $\phi$ are pre-computed and stored. Their values are interpolated for intermediate values of Reynolds number and volume fraction. This allows rapid evaluation of the PIEP force and torque based on these maps.

In an EL simulation, such as that shown in \ref{fig9}, the torque on the $i^{th}$ particle is calculated as the sum of the undisturbed flow and quasi-steady contributions as \cite{akiki2017pairwiseJCP}
\begin{equation}\label{eq:32}
\bT_i = \bT_{un,i} + \bT_{qs,i} \, .
\end{equation}
The undisturbed flow torque contribution can be expressed as
\begin{equation}\label{eq:33}
\bT_{un,i} = \dfrac{\pi}{2 \Rey} \left[ \overline{ \left( \left[ \bn \cdot \dfrac{(\nabla \bu_{0} + \nabla \bu_{0}^T)}{2}  \right] \times \bn \right)}^S_i + \sum_{j=1, j\ne i}^{N} |\bu_{r,j}|^2 \mathscr{M}_{T2} \, \bbe_o \right] \, .
\end{equation}
Similarly the quasi-steady torque contribution can be expressed as
\begin{equation}\label{eq:34}
\bT_{qs,i} = \dfrac{\pi}{\Rey} \left[ \dfrac{1}{2} f_\omega \left( \overline{ (\nabla \times \bu_0)}^V_i + \sum_{j=1, j\ne i}^{N} |\bu_{r,j}| \mathscr{M}_{T1} \bbe_o \right) - f_\Omega \bOmega_i \, \right] \, ,
\end{equation}
where the finite Reynolds number correction factors $f_\omega$ and $f_\Omega$ are presented in \cite{akiki2017pairwiseJCP, bagchi_bala1, bagchi_bala2}.

In the EL simulation, along with the translational motion of the particles, their rotational motion is also solved. The rotational motion of the particles is driven by the hydrodynamic torque exerted on them by the surrounding fluid as given by the sum of the undisturbed flow and quasi-steady contributions given in \eqref{eq:32}. In addition, collisional interactions contribute to additional torque on the particles as implemented in the soft-sphere collisional model. 

\section{Conclusions}
The primary objective of the present work is to develop a modeling approach that provides particle-resolved-like accuracy to Euler-Lagrange simulations. In Euler-Lagrange simulations the continuous phase governing equations are filtered (or averaged) and only length scales that are much larger than the particle size are resolved. As a result, the momentum exchange between the particles and the surrounding fluid flow must be accounted for in terms of force and torque models. The standard models that are dependent on $\Rey$ and $\phi$ are accurate in capturing the mean force exchange between the particles and the surrounding fluid, but do not account for the strong particle-to-particle variations that arise primarily due to the unique arrangement of each particle with respect to its neighbors and the perturbation flow induced by them. Thus, in order to obtain particle-resolved-like accuracy, each particle must account for the relative location of its nearest neighbors and the force on the particle must be modeled taking into account the perturbation effect of the neighbors.

We pursued two different machine learning approaches to develop improved force models that takes into account the effect of neighbors: (i) a direct approach to force prediction that is based on artificial neural network (ANN) and (ii) an indirect two-step approach that uses flow prediction followed by force prediction. The accuracy of the machine learning algorithms depends on the quantity of training data provided to the training process. The direct ANN approach requires large amounts of PR-DNS training data of drag forces on particles sampled over a wide range of arrangement of neighbors, in terms of their relative position, velocity and acceleration. Without such extensive data, the force prediction by ANN suffers from over-fitting. The alternative two-step approach simplifies the problem by invoking the pairwise interaction approximation. In the first {\it {flow prediction}} step, a linear regression algorithm is used with the PR-DNS data of steady flow over a random array of frozen particles to obtain the optimal estimation of the perturbation flow due to a particle as the superposable wake. In the second {\it {force prediction}} step, the perturbation force on a particle from each of its neighbors' perturbation flow, taken one at a time, is evaluated using the generalized Fax\'en form of the Maxey--Riley equation. This estimation of perturbation force was then corrected with additional force correction maps that were obtained from PR-DNS data using linear regression. The two-step process results in the pairwise interaction extended point-particle (PIEP) force model, which has also been extended to torque prediction.

It is observed that the two-step process of flow prediction followed by force prediction is crucial for a successful implementation of the machine learning algorithm. This is because the PR-DNS data of force and torque information is available for only a few thousand particles, however, the flow field around these particles involves tens of millions of grid points. Thus, far more training data is available for flow prediction than for direct force prediction. Furthermore, we have developed flow prediction in the form of superposable wakes for a wide range of particle Reynolds number and volume fraction using machine learning in the context of a stationary random array of particles. However, subsequent application of force prediction using Generalized Fax\'en law is in the dynamic context of freely moving distribution of particles. Thus, our strategy is to use machine learning algorithms for accurate flow prediction followed by multiphase theory for dynamic force prediction. The above hybridization of multiphase physics and machine learning is particularly important, since it blends the strength of each, and the resulting force model cannot be developed by either physics or machine learning alone. While the physics provides the decomposition of force into components and their functional forms, machine learning extracts the detailed perturbation influence of the neighbor.

The performance of the PIEP and the ANN models were evaluated by first applying them to predict the force on a random array of stationary particles subjected to a steady uniform flow. This configuration was also studied using a particle-resolved simulation, which served as an independent test data that was entirely different from of those used for training the models. The $R^2$ values for the ANN prediction were negative, and therefore were worse than the mean drag model for all combinations  $\Rey$ and $\phi$, indicating over-fitting of the training data. On the other hand, the PIEP model retains its predictive capability and was able to capture approximately 53\% to 75\% of the variations in particle forces depending on $\Rey$ and $\phi$. This should not be interpreted as a fundamental weakness of ANN's ability to arrive at an accurate model. This merely indicates the need for much larger amount of PR-DNS data if we want to pursue direct force prediction. However, machine learning for flow prediction followed by physics-based force prediction is a viable strategy in the face of limited data. 

We also considered the application of PIEP model for the dynamic case of 168972 sedimenting particles in a triply periodic box. Here we demonstrated the ability of the superposable wake to predict the microscale flow around the distribution of particles. With a superposition of perturbation flow around all the particles within the system, one could obtain particle-resolved-like subgrid accuracy in an EL simulation from the knowledge of the particles and their motion. Since a PR-DNS of this sedimentation problem is not possible, we compared the solution of EL simulations with and without the PIEP model. The comparison metric chosen was the rate of inter-particle collisions that occur in the system. With fluid-mediated inter-particle interactions being evaluated with the PIEP model, the number of particle collisions per time step is nearly doubled, indicating the importance of accounting for neighbors' influence in the force and torque models.

Although the proposed hybrid-PIEP model provides substantial improvement over the standard drag correlation, there are significant limitations. The foremost among them is the assumption of pairwise interaction used both in the evaluation of the undisturbed flow in the flow prediction step and in the use of the Generalized Fax\'en law in the force prediction step. Another important limitation is the use of perturbation maps trained under static condition being deployed under dynamic conditions. This quasi-steady approximation adequate under conditions of small acceleration, since the hybrid PIEP model accounts for unsteady effects with the inclusion of the stress-divergence and added mass contributions to force. However, under conditions of strong acceleration, either of the particle or of the surrounding flow as in the case of shock propagation, the perturbation flow due to a neighbor will also include a dominant inviscid potential flow contribution due the acceleration, whose effect must also be considered. 

	\section{Acknowledgments}
	This material is based upon work supported by the U.S. Department of Energy, National Nuclear Security Administration, Advanced Simulation and Computing Program, as a Cooperative Agreement under the Predictive Science Academic Alliance Program, under Contract No. DE-NA0002378, by the Office of Naval Research (ONR) as part of the Multidisciplinary University Research Initiatives (MURI) Program, under grant number N00014-16-1-2617 and by the National Science Foundation Graduate Research Fellowship Program under Grant No. DGE-1315138 and DGE-1842473.
	
\bibliographystyle{aiaa}
\bibliography{bibfile}

\end{document}